# Decoupling Control From Data for TCP Congestion Control

A thesis presented

by

Shie-Yuan Wang

to

The Division of Engineering and Applied Science

in partial fulfillment of the requirements
for the degree of

Doctor of Philosophy

in the subject of

Computer Science

Harvard University

Cambridge Massachusetts

September 1999

# Abstract

Many applications want to use TCP congestion control to regulate the transmission rate of a data packet stream. A natural way to achieve this goal is to transport the data packet stream on a TCP connection. However, because TCP implements both congestion and error control, transporting a data packet stream directly using a TCP connection forces the data packet stream to be subject to TCP's other properties caused by TCP error control, which may be inappropriate for these applications.

The ***TCP decoupling*** approach proposed in this thesis is a novel way of applying TCP congestion control to a data packet stream without actually transporting the data packet stream on a TCP connection. Instead, a TCP connection using the same network path as the data packet stream is set up separately and the transmission rate of the data packet stream is then associated with that of the TCP packets. Since the transmission rate of these TCP packets is under TCP congestion control, so is that of the data packet stream. Furthermore, since the data packet stream is not transported on a TCP connection, the regulated data packet stream is not subject to TCP error control.

Because of this flexibility, the TCP decoupling approach opens up many new opportunities, solves old problems, and improves the performance of some existing applications. All of these advantages will be demonstrated in the thesis.

This thesis presents the design, implementation, and analysis of the TCP decoupling approach, and its successful applications in TCP trunking, wireless communication, and multimedia streaming.

# Acknowledgments

I would like to thank my advisor, Professor H.T. Kung, for the advice, support, and care that he gave me during my stay at Harvard. Without his support and care, the journey would have been harder and longer. Professor Michael Smith, Dr. Alan Chapman of Nortel, Sally Floyd of ACIRI, and Ramachandran Ramjee of Lucent all gave me many valuable comments on my Ph.D. thesis. I would like to thank them all.

My other thanks go to my friends Robert Morris and Brad Karp. I learned a lot of knowledge and received a lot of assistance from them. I wish that I could have met them earlier in my research life.

During my stay at Harvard University for pursuing my Ph.D. degree (1995 Fall to 1999 Fall), my parents in Taiwan underwent the 1996 Taiwan missile crisis, the 1997 Asia economy crisis, and the 1999 Taiwan earthquake (which happened two days before my Ph.D. thesis defense and killed more than 2,000 people). If it had not been for their hard work and the sacrifices that they made which allowed them to encourage and support me, I would have had to quit my Ph.D. program. My wife Mei-Lin discontinued her Ph.D. program at Boston University so that she could care for me, my daughter, and my son. She also returned to Taiwan with my daughter and son in the summer of 1998 to help my parents earn money to support me. Without her sacrifice, I could not receive a Harvard Ph.D. degree and at the same time have a happy family.

I owe too much to my parents and my wife. I want to dedicate my Ph.D. thesis to my parents and wife. The journey at Harvard has been fruitful and unforgettable. I hope that in the future my daughter and son also can luckily have a chance to study at Harvard like their father.

<div style="text-align: right">September 22, 1999</div>

# *Table of Contents*

# Chapter 1  Introduction

TCP [32, 62] is the main congestion and error control method for the Internet. While being able to provide reliable transmission between the two end hosts of a connection, TCP can control the connection's bandwidth usage to avoid network congestion [59, 63]. Over the years, researchers have built up a large body of knowledge about TCP [33, 42, 40, 34, 53, 54, 48, 43, 23, 44, 55, 19], regarding its throughput, fairness, robustness, and its interactions with various packet scheduling and buffer management schemes employed in routers.

Although TCP's design of combined error and congestion controls has been successful for "traditional" applications such as telnet, ftp, email, and WWW for a long time, it has been difficult to use TCP in some new application areas where independent uses of TCP's congestion control and TCP's error control is desired. Examples of such applications include TCP trunking [24], transport over wireless links with high bit-error rates, and transport of real-time, multimedia data.

An elastic TCP trunk [24] is a trunk whose transmission rate is controlled by TCP congestion control. Since a trunk should function like a logical link that does not automatically retransmit lost packets, a TCP trunk should use TCP's congestion control to probe for and use available bandwidth, but should not use TCP's error control to endlessly retransmit a lost data packet until the data packet eventually reaches the other end of the trunk. Reliable transport service over a lossy wireless link benefits from TCP's error control to retransmit packets corrupted by bit errors on wireless links and TCP's congestion control to avoid congestion, but must avoid invoking TCP congestion control to mistakenly reduce the current sending rate upon packet losses caused by link errors. Real-time UDP packet streams, such as those for audio and video, are delay- and jitter-sensitive. TCP's congestion control should adapt the transmission rate of these audio/video streams to the available bandwidth, but TCP's error control should not be invoked, because a time-sensitive packet which is lost and automatically retransmitted may still be too late to be useful for the audio/video application. The retransmission decision for such time-sensitive packets should be left to the application.

This thesis proposes an approach, called the ***TCP decoupling***, for solving or alleviating the above difficulties encountered when applying TCP congestion control to TCP trunking, wireless communication, and multimedia streaming applications. With this approach, TCP's congestion control alone can be applied to a data packet stream (can be a single flow or an aggregate flow) without imposing TCP's error control onto it.

This thesis considers three applications which benefit from using the TCP decoupling approach. The first application creates a new kind of trunk, called TCP trunk, which uses TCP congestion control to probe for and use available bandwidth while keeping packet drop rates low in network. The second application provides a reliable transport-layer service on hosts that manages TCP's congestion control independently from TCP's error control. In this case, the TCP decoupling approach improves a TCP connection's throughput on lossy wireless links by a factor that is proportional to sqrt(**MTU**/**HP_Sz**), where MTU is the wireless link's maximum transmission unit and HP_Sz the size of a packet containing only a TCP/IP header. For example, when MTU is 1500 bytes (the MTU of the wireless Wavelan [65] network) and HP_sz is 40 bytes, this factor is 350%. The third application provides an unreliable transport-layer streaming service on hosts that applies TCP's congestion control to a UDP packet stream so that the UDP packet stream becomes 100% TCP-friendly when competing for available bandwidth with TCP connections. Design and implementation of these applications in UNIX kernels (FreeBSD 2.2.7) and experiments on laboratory testbed networks will be presented. This thesis will demonstrate that the TCP decoupling approach leads to improved performance and simple and efficient implementation for these applications.

The important contributions of this thesis are as follows. First, this thesis proposes the TCP decoupling approach and shows that it is a new, general, and powerful method for implementing congestion control in networks. Second, specifically and concretely, this thesis shows that the TCP decoupling approach improves performance and creates new useful services in TCP trunking, wireless communication, and multimedia streaming applications.

This thesis is organized as follows. Chapter 2 presents the TCP decoupling approach. Chapter 3, 4, and 5 presents the application of the TCP decoupling approach in TCP trunking, wireless communication, and multimedia streaming, respectively. Finally, Chapter 6 concludes this thesis.

# Chapter 2  The TCP Decoupling Approach

## 2.1 Overview of TCP

This section briefly overviews TCP as TCP is the basic of the TCP decoupling approach. This thesis does not intend to modify and improve TCP congestion control. Instead, the contribution of this thesis is to decouple TCP's error control from TCP's congestion control, and to apply the best available TCP congestion control to some applications to improve performances or create new useful services.

TCP provides a reliable and in-sequence transport and delivery service for a pair of application sender and application receiver running at the user level. To use TCP to transport data between the application sender and receiver, a TCP connection needs to be set up between them. A TCP connection is represented by its two end points at the sending and receiving hosts where the application sender and receiver are running. These end-points, called "socket" and containing the values of many TCP state variables for a TCP connection, are stored in a TCP control block in the kernel to identify a particular TCP connection. The application sender generates its data and write the data to the socket of its TCP connection. Data in the socket is viewed and treated as a stream of bytes before being transported on the TCP connection. The TCP processing module in the kernel of the sending host (called "TCP sender" in the following discussion for the sake of simplicity) then segments the stream of data bytes into multiple segments each of MSS (maximum segment size) bytes, encapsulates each of them in a TCP packet with a TCP/IP header, and then send these data packets to the receiving host. The TCP processing module in the kernel of the receiving host (called "TCP receiver" in the following discussion for the sake of simplicity), after receiving these data packets, will generate acknowledgment packets and send them to the TCP sender to acknowledge the receipt of the transported data. The TCP receiver will then deliver the received data to the socket of the TCP connection, from where the application receiver can read them.

To provide a reliable service, a lost data packet is retransmitted until the application receiver finally receives and acknowledges it. To provide an in-sequence delivery service, if a data packet (and the data carried in it) is lost, the data carried in the lost packet's following out-of-order data packets will not be immediately delivered to the application receiver when these data packets arrive at the receiving host. Instead, these out-of-order data packets will be queued in an assembly queue in the kernel of the receiving host until the lost data packet is retransmitted and finally arrives. At that time, consecutive and in-sequence data will be delivered to the application receiver. For identifying which segment of the data byte stream is being sent and reporting which segments have been received, a "sequence number" and an "acknowledge number" field [62] are used in a TCP header, respectively. The unit of these numbers are byte. Figure 1 shows the format of a TCP/IP packet.

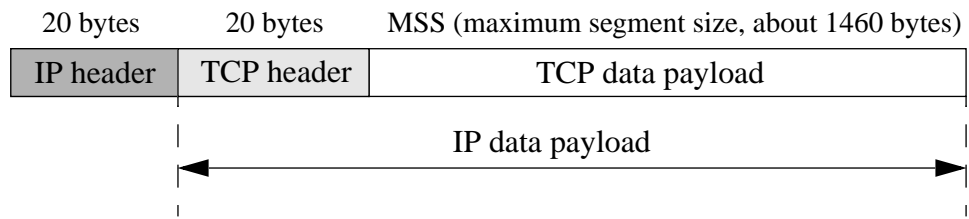

Figure 1. The format of a TCP/IP packet. TCP's MSS (maximum segment size) is normally about 1460 bytes in Ethernet networks. It is derived from deducting the sizes of IP and TCP headers from the Ethernet MTU (maximum transmission unit), which is 1500 bytes.

TCP's error control uses a cumulative acknowledgment scheme to acknowledge consecutive data bytes that have been received and delivered to the application receiver. For each TCP connection, the value of the variable rcv_nxt is kept in the TCP connection's control block. This value will be updated when a data packet arrives to reflect that all the bytes between the sequence number 0 and rcv_nxt have been correctly received and delivered to the application receiver. To save network bandwidth, when there is no data packet loss, the TCP receiver sends back an acknowledgment packet containing the current value of rcv_nxt for every two data packets. If a data packet is lost, the arrival of its following out-of-order data packets will not change the current value of rcv_nxt. Instead, for each arrived out-of-order data packet, an acknowledgment packet containing the current value of rcv_nxt will be sent back to the TCP sender. These acknowledgment packets with the same acknowledgment numbers are called "duplicate acknowledgments" in TCP.

TCP's congestion control is complex and contains four intertwined [64] algorithms: slow start, congestion avoidance, fast retransmit, and fast recovery. Since taking advantage of TCP's congestion control, rather than refining TCP's congestion control, is the focus of the thesis, this section will only give a high-level description of TCP's congestion control and will not present these four algorithms in detail.

TCP's congestion control is a window-based congestion control scheme. It dynamically restricts the sending rate of a TCP connection by restricting the TCP connection's maximum number of outstanding packets in the network. An outstanding packet is a packet that is already sent by the TCP sender but its corresponding acknowledgment packet has not been received by the TCP sender. The number of allowable outstanding packets is called "congestion window size" in TCP and its value is stored in the variable "cwnd" [20] in the TCP connection's control block at the sending host. (In TCP implementation, the unit of TCP congestion window size is in bytes, not in packets. Because normally each packet of a TCP connection is of the same size, for the sake of simplicity, this thesis uses packet instead of byte as the unit of TCP congestion window size.) Since dividing the congestion window size of a TCP connection by the TCP connection's round-trip time (RTT) is the TCP connection's sending rate (also called "throughput"), and normally the RTT of a TCP connection remains about the same during its life time, adjusting the congestion window size of a TCP connection achieves the goal of controlling a TCP connection's sending rate.

After a TCP connection is set up, if the application sender always has data to send (this kind of TCP connection is called "greedy" TCP connection), the TCP sender will double its congestion window size (and thus its sending rate) every the TCP connection's

RTT until some of its data packets get lost. This algorithm is called "slow start" in TCP and is used to quickly probe the available bandwidth.

When a data packet gets lost, the TCP receiver will send duplicate acknowledgment packets back to the TCP sender for each out-of-order data packets (explained above when presenting TCP error control). After having received three duplicate acknowledgment packets, the TCP sender thinks that a data packet is lost and treats the packet loss as a signal of congestion. The TCP sender immediately retransmits the lost packet and then shrinks its current congestion window size by a half, which results in a 50% sending rate reduction to mitigate congestion. This algorithm is called "fast retransmit" and is used to quickly reduce the current sending rate and retransmit a lost packet. After the first fast retransmit, the TCP sender starts to enter the "congestion avoidance" phase in which the TCP sender gradually increases its congestion window size by one packet every the TCP connection's RTT to probe for and utilize available bandwidth if no more packets are lost. In case a packet is lost again and three duplicate acknowledgment packets have been received, the fast retransmit will occur again, and the additive-increase and multiplicative-decrease congestion control algorithm repeats. Because the additive-increase and multiplicative-decrease congestion control principle is effective [7], although many TCP variants such as TCP Tahoe [40], TCP Reno [40], TCP SACK [43], etc. have been proposed, all of them still use this additive-increase and multiplicative-decrease congestion control principle and differ only in minor modifications to fast retransmit and recovery algorithms.

A TCP connection has a non-zero inherent packet loss rate even when the network is uncongested. The reason is that, during either the slow start or congestion avoidance phase, a TCP connection needs to periodically lose some packets in order to probe for available bandwidth in the network.

If multiple packets in a congestion window are dropped, in TCP Tahoe and Reno, the TCP sender may time-out and halt the data transfer for at least one second. The reason is that for every lost data packet, fast retransmit will be triggered and the congestion window size be reduced by one half again. After several reductions, when the congestion window size becomes smaller than three packets, there will not be enough three duplicate acknowledgment packets to trigger fast retransmit. As a result, the TCP sender has to time-out for at least one second. After timing-out, the TCP sender will restart its data transfer using the slow start algorithm. Recently, TCP SACK [43] has been proposed to mitigate the time-out problem. In [40, 49], experimental results show that TCP SACK is able to recover from multiple losses within one window of data without necessarily timing-out.

To implement TCP's congestion and error control, some control information is stored in a packet's TCP header [62]. The information includes the source and destination port numbers for identifying the endpoint of a TCP connection on a host, checksum for detecting corrupted data (for error control), sending sequence numbers for identifying the data being transmitted, acknowledge sequence number for acknowledging received data, advertised window size for flow control, and various control flags for identifying special control packets such as SYN, FIN packets transmitted when a TCP connection is being set up or torn down. For a TCP connection, at its sending and receiving hosts, the TCP processing module keeps the values of many TCP state variables (e.g., snd_nxt, rcv_nxt, cwnd [20]) and store them in a TCP control block. The sending and receiving hosts update and exchange these information via packets to work together to implement TCP's error and congestion control in an end-to-end way.

In the current TCP design, TCP's error control and TCP's congestion control are coupled together and cannot be used independently. When a data packet is lost, both TCP error and TCP congestion controls take actions: TCP error control retransmits the lost data packet and TCP congestion control reduces the congestion window size. This coupling causes some problems as follows. First, TCP's congestion control may be unnecessarily and mistakenly affected by TCP's error control. For the case of packet dropping due to buffer overflow, the rate reduction action taken by TCP's congestion control is correct. However, for the case of packet corruption due to link errors, the rate reduction action taken by TCP's congestion control is unnecessary and only leads to poor throughput [35] on error-prone links such as wireless links. Second, TCP's congestion control cannot be applied alone to some applications without TCP's error control. The reason is that, by its service definition [32], TCP insists on providing a reliable and in-sequence delivery of data at the receiving host. As a result, if a lost packet is not retransmitted by the sending host and eventually received by the receiving host, the data transfer will halt. The reasons are as follows. First, TCP's congestion control implements a flow control to prevent buffer from overflowing at the receiving host. Second, the receiving host has only a limited buffer space to store out-of-order packets (those packets sent after the lost packet at the sending host). Third, the receiving host cannot deliver and release the buffer space occupied by these out-of-order packets until the lost packet finally arrives.

Due to the inflexibility of coupling TCP error and congestion control, a TCP connection achieves poor throughput in wireless networks and is not suitable for transporting a data packet stream which needs only TCP congestion control but not TCP error control. Examples of such a data packet stream includes a real-time audio/video UDP packet stream or a trunk packet stream.

## 2.2 Design and Implementation of the TCP Decoupling Approach

### 2.2.1 Notations and Terminologies

For the sake of simplicity and accuracy, this section defines some notations and terminologies which will be used throughout the thesis.

**Circuit**

A routing path over which a stream of data packets will be transmitted. Data packets to enter a circuit will be emitted into the circuit at certain rates by the sending node of the circuit. Once a data packet is emitted into the circuit, it will be forwarded by intermediate routers on the circuit path as soon as they can.

**TCP Circuit**

A circuit whose sending node uses TCP congestion control to control the emission rate of data packets into the circuit.

**Control TCP**

A control TCP is a TCP connection set up between the sending and receiving node of a TCP circuit to regulate the emission rate of a data packet stream flowing into the circuit. The version of control TCP used in the current implementation of TCP decoupling is TCP-Reno. It can be any better version (e.g., TCP-SACK) when it is proposed in the future.

**Sender and Receiver of a Control TCP**

The sender of a control TCP refers to the TCP processing module at the sending node of the control TCP. The receiver of a control TCP refers to the TCP processing module at the receiving node of the control TCP.

**Header Packets**

These are packets generated and sent by the sender of a control TCP to the receiver of the control TCP. They contain only TCP/IP headers, that is, they have empty TCP data payloads.

**Control Packets**

Control packet are either header packets or the acknowledgment packets generated and sent back by the receiver of the control TCP to the sender of the control TCP.

**Data Packets (User Packets)**

Data packets are those packets which are not control packets in a network. Since a data packet is normally generated by a user application program, it is also called a "user" packet in the thesis.

**GMB**

 Guaranteed Minimum Bandwidth, in bytes per unit time.

**VMSS**

VMSS stands for "virtual maximum segment size". It is a configurable constant.

**α**

    Required size of the TCP congestion window, in packets, for fast retransmit and recovery to work well. This paper assumes $\alpha = 8$.

**β**

    The fraction of a link bandwidth allocated for GMB traffic. This paper assumes $\beta < 1$.

**N**

    Number of TCP connections sharing a router buffer.

**HP_Sz**

    Header packet size in bytes. This paper assumes $HP\_Sz = 52$ because typically a header packet contains a 40-byte TCP/IP header and a 12-byte TCP timestamp option.

**HP_Th**

    Header packet threshold in header packets. A router on the path of a control TCP will drop arriving header packets, when their number in the router buffer exceeds HP_Th.

### 2.2.2 Overview of the TCP Decoupling Approach

The TCP decoupling approach is developed to implement a TCP circuit. A TCP circuit is a circuit which applies TCP's congestion control to a data packet stream when it flows through the circuit without imposing TCP's error control onto it. Figure 2 (a) depicts a TCP circuit. A TCP circuit can be used as an edge-to-edge TCP trunk (to be presented in Chapter 3) or an end-to-end connection for wireless communication and multimedia streaming (to be presented in Chapter 4 and 5, respectively). The sending node and receiving node of a TCP circuit thus can be a router or a host, depending on where the TCP decoupling approach is used. For TCP trunking application where a TCP trunk aggregates traffic from different sources before the traffic enters the core network, because a TCP trunk starts and ends at an edge router (this is an edge-to-edge application), the sending and receiving node each is a router. For wireless communication and multimedia streaming applications, because the application programs (the traffic generator and sinker) run on hosts (they are end-to-end applications), the sending and receiving node each is a host. The same TCP decoupling approach can work for both edge-to-edge and end-to-end applications without any change.

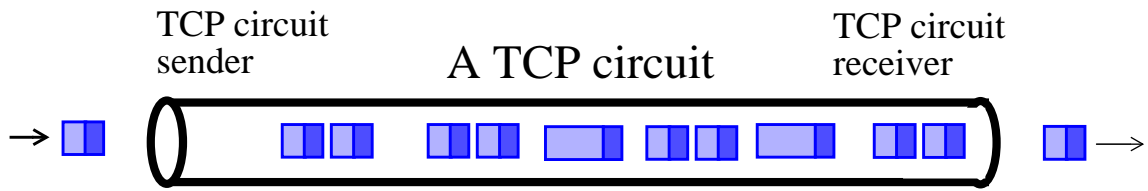

(a) A TCP circuit

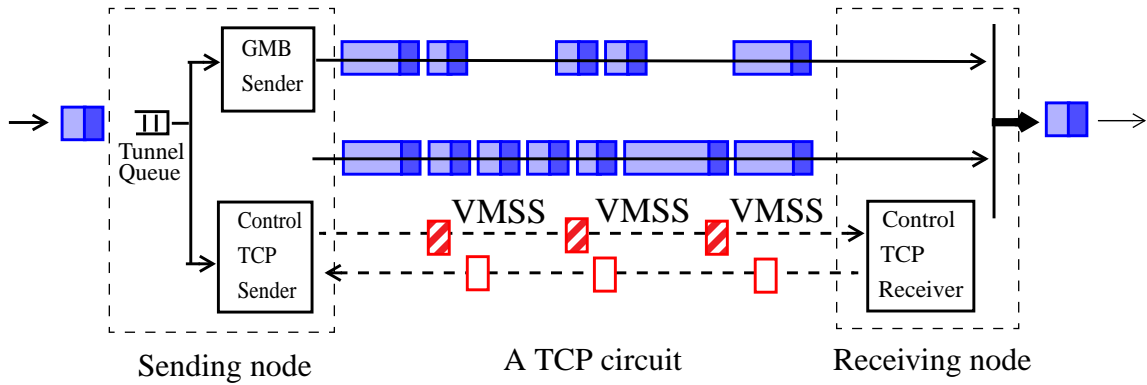

(b) The TCP decoupling implementation for a TCP circuit

Figure 2. The TCP decoupling approach. for implementing a TCP circuit.

Figure 2 (b) shows how a TCP circuit is internally implemented by the TCP decoupling approach. A TCP circuit is composed of a GMB sender at its sending node and one or multiple control TCPs between its sending and receiving nodes. Data packets flowing into the TCP circuit are first stored in the tunnel queue at the sending node. The GMB sender is used when a TCP circuit is allocated a certain GMB along its path. The GMB sender unconditionally sends data packets in the tunnel queue into the TCP circuit at the GMB rate. In Figure 2 (b), one TCP connection is set up between the sending node and the receiving node of the TCP circuit to probe for the available bandwidth for the data packet stream beyond the TCP circuit's allocated GMB.

The TCP connection, which is called "control TCP", sends out its header packets under TCP congestion control algorithms when there are data packets in the tunnel queue. These header packets all contain only a TCP/IP header and no data payload as in order to implement and use TCP congestion control, the control information exchanged and carried by the packets of a TCP connection is all contained in the TCP headers of these packets, and the content of the TCP data payloads of these packets are totally irrelevant and can be empty. For each transmitted header packet, the control TCP on the sending node emits data packets in the tunnel queue into the TCP circuit totalling up to VMSS bytes. The sending rate of the data packet stream thus is proportional to the sending rate of the header packets. Since data packets traverse the same routing path as header packets (this assumption is discussed in Section 2.2.3), they will experience the same congestion level at the same place at the same time. Suppose that congestion occurs and buffer eventually overflows in a router on the path, which results in dropping of header packets, the sender of the control TCP will reduce the sending rate of its header packets, which results in a proportional reduction in the sending rate of the data packet stream. By this method, the TCP decoupling approach achieves the goal of using TCP's congestion control to regulate the transmission rate of a data packet stream for utilizing available bandwidth.

In contrast with the traditional TCP approach in which data packets need to be carried (encapsulated) by TCP packets and thus be coupled with TCP/IP headers, in the TCP decoupling approach, data packets are transmitted as independent packets from control packets and their packet format and packet content remain unchanged. The data packet stream does not suffer from the problems caused by TCP's error control as TCP's error control is applied to the header packets only, not to the data packets.

### 2.2.3 Assumption of the TCP Decoupling Approach

One assumption required by the TCP decoupling approach is that the routing path taken by the data packets should be the same as the routing path taken by the header packets which control them. Obviously, if header packets take a different path than data packets, header and data packets will not experience the same congestion level at the same place at the same time in a network and, as a result, the TCP decoupling approach may fail.

Fortunately, in the current Internet, it can be argued that when the TCP decoupling approach is used for end-to-end applications, the problem of using different routing paths for header and data packets is not likely to happen, and when the TCP decoupling approach is used for edge-to-edge applications, there exist solutions for it. The reasons are presented as follows.

First, for end-to-end applications such as wireless communication and multimedia streaming, the IP destination addresses of the data and header packets are the same. When routing tables in routers do not change, both of them will take the same routing path. When some routing tables suddenly change, a header packet may take a different routing path than its associated data packet(s). Although this problem may happen, it affects only one pair of a header and its associated data packets totaling up to VMSS bytes. The route change problem also happens infrequently as the Internet routing protocol OSPF [5, 50] only updates routers' routing tables every 30 seconds to avoid route flapping. Another concern is about multi-path routing, which splits the load of a packet stream onto many routing paths for load-balancing purposes. This is not a concern as network researchers now understand that the minimum granularity of load-balancing should be a flow -- a packet stream with the same source and destination IP addresses, otherwise a TCP connection's throughput and the quality of a UDP audio/video stream will suffer a lot from excessive packet reordering. (For applications using TCP, packet reordering causes duplicate acknowledgment packets, which unnecessarily trigger TCP's fast retransmit algorithm, which in turn unnecessarily reduces the sending rate of a TCP connection. For multimedia applications using UDP, packet reordering increases the required buffer size to store and rearrange out-of-order packets before they can be played back at the receiving host, which also adds unnecessary delays to the playback time and degrades the quality of real-time applications such as IP phone).

Second, for the edge-to-edge application such as TCP trunking where the data packet stream is composed of many flows each with its own different IP source and destination addresses, although the above arguments for end-to-end applications fail to hold, a TCP trunk can be associated with a layer-2 ATM [8] or Frame Relay [21] virtual circuit, or an MPLS label-switched path [51, 9] to make sure that its header and data packets all take the same routing path. Since TCP trunks are intended to be used in the backbone networks (see Chapter 3), where ATM and Frame Relay virtual circuits are being used extensively and MPLS is designed for engineering traffic, running a TCP trunk on top of a virtual circuit or a label-switched path is feasible and well-suited.

### 2.2.4 Design Goals (Also Properties) of the TCP Decoupling Approach

The design goals of the TCP decoupling approach are listed as follows:

1. Data packets trigger control (header) packets

2. Do not automatically retransmit lost data packets

3. Do not introduce packet reordering to a data packet stream

4. Do not introduce extra transmission delay to a data packet other than that caused by TCP's congestion control

5. Do not modify the content of a data packet

6. Do not increase the length of a data packet

7. Low bandwidth overhead for control packets

8. Simple and efficient implementation (high throughput)

9. Easy to set up, configure, and use

Goal (1) is desirable because sending control packets when there are no data packets to send is unnecessary and wastes network bandwidth. Goal (2) is desirable because different applications have different reliability requirements for their packet transfer (e.g., FTP requires reliable data transfer and video-conferencing can tolerate unreliable transfer), retransmitting data packets, if required, should be handled by the application program or some reliable protocol at the sending host. Goal (5) is desirable because modifying a data packet's content needs several read/write operations and recomputation of the IP checksum, which will slow down the forwarding throughput. Goal (6) is desirable because increasing a packet's length may cause packet fragmentation when the resulting length exceeds the MTU (maximum transmission unit) of some link on which the packet need to traverse. Goal (7) is desirable because the overhead control packets should not consume too much bandwidth. Goal (8) is desirable because a simple implementation leads to a low-cost and robust implementation, and an efficient implementation can provide high forwarding throughput on high speed link such as OC-12 (622 Mbps) and OC-48 (2.4 Gbps).

The design and implementation of the TCP decoupling approach meets all of these goals. Therefore, these listed design goals are also the general properties of the TCP decoupling approach. In addition, the TCP decoupling approach has many other properties that are specifically useful in TCP trunking, wireless communication, and multimedia streaming applications. These application-specific properties will be presented in Chapter 3, 4, and 5, respectively.

## 2.2.5 The TCP Decoupling Mechanism on the Sending Node of a TCP Circuit

Figure 3 depicts the architecture of the sending node of a TCP circuit in the TCP decoupling approach. Each component will be presented in detail in the following sections.

### 2.2.5.1 Tunnel Queue

Data packets that are to be sent into a TCP circuit are first redirected to and enqueued in a tunnel network interface queue from which they will later be dequeued and forwarded by either the GMB sender or one control TCP sender. A tunnel network inter-face is a pseudo network interface that does not have a real physical network attached to it [17]. Its functions, however, from the kernel's point of view, are no different from those of a normal Ethernet network interface. The tunnel interface queue serves as an input queue for temporarily holding data packets not yet forwarded out. Although using any software queue in the kernel also works for serving as an input queue, using a tunnel network interface queue has an advantage. The advantage is that, since from the kernel's point of view, a tunnel network interface is like a physical network interface, redirecting arriving data packets to a tunnel interface queue can be done simply by changing just one routing entry in the sending node's routing table. When allowed by the GMB rate or TCP congestion control, the GMB sender or one control TCP sender dequeues the first redirected data packet and calls the kernel's IP packet forward function (ip_forward()) to forward it out. Figure 4 depicts the data packet redirection scenario.

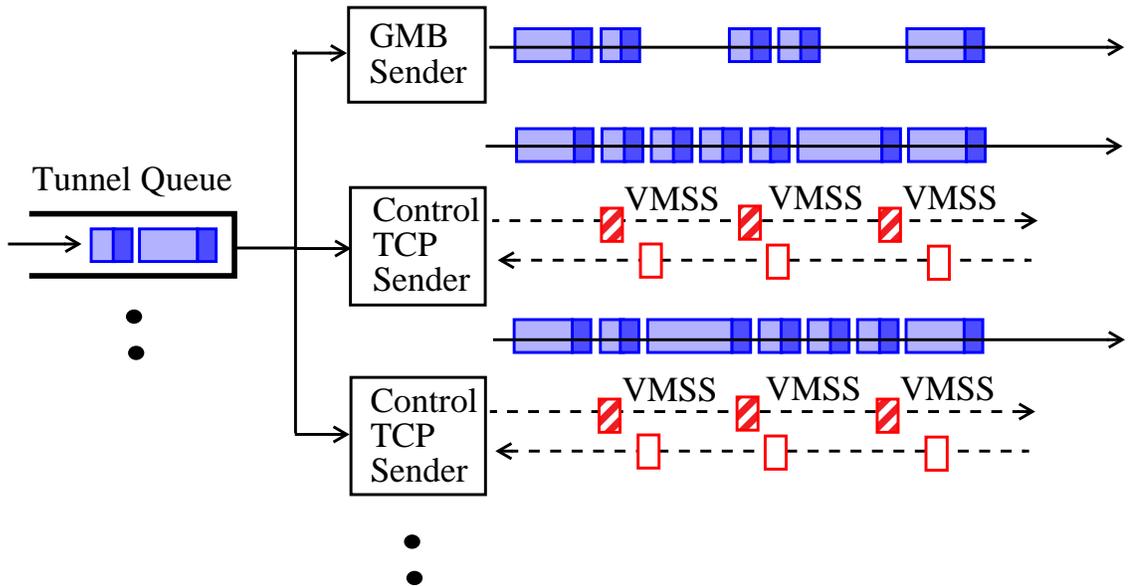

Figure 3. The architecture of the sending node of a TCP circuit implemented in the TCP decoupling approach.

Redirecting arriving data packets to a tunnel interface queue by changing a routing entry for them results in a routing loop problem. The problem is that the redirected data packets, when they are dequeued and forwarded by ip_forward(), will be forwarded back to the tunnel interface and enqueued there again. What causes this problem is that ip_forward(), when given a packet to forward, first looks in the routing table to retrieve the correct routing entry for the outgoing packet and then passes the retrieved routing entry as an argument to ip_output(), which then sends the packet to the network interface specified in the routing entry. Since a routing entry has been changed so that packets with a particular IP destination address are redirected to the tunnel interface, the redirected packets will be forwarded back and redirected forever. This looping problem is solved as follows. First, a routing entry for a special unused IP destination address (e.g., 99.99.99.99) is created. This routing entry is configured to point to the correct physical network interface (it is the Ethernet interface in Figure 4) for the already-redirected data packets. Second, instead of calling ip_forward(), the GMB sender or the control TCP sender directly calls ip_output() with this special routing entry as one argument.

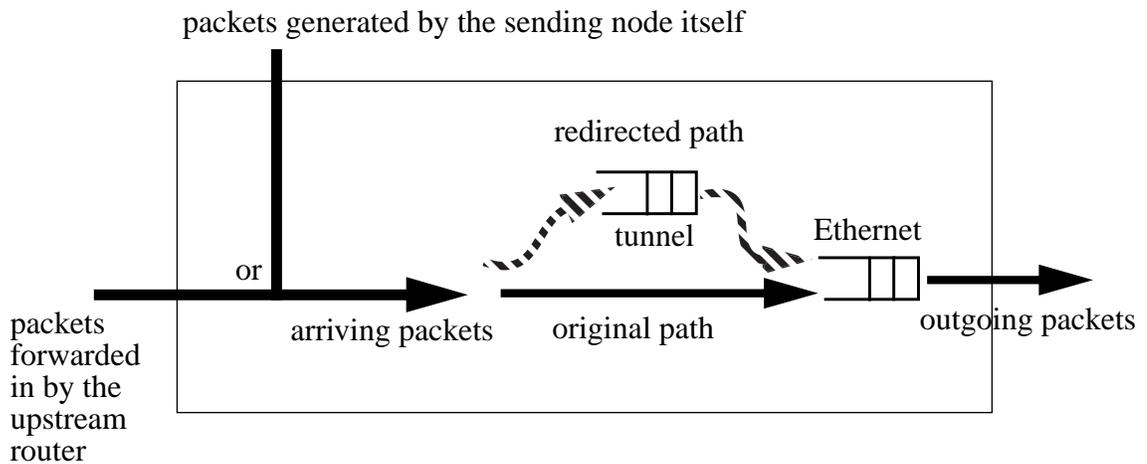

Figure 4. Arriving packets are redirected to and enqueued in a tunnel network interface queue. They will be sent out through a physical network interface (e.g., an Ethernet interface) later.

Multiple queues can be used in the tunnel network interface to provide a differentiated service for packets transported in the data packet stream if their application programs mark them with different priorities. An absolute priority or weighted round-robin [2] scheduling scheme can be used to allocate the achieved bandwidth of the TCP circuit among different classes of packets that compose the data packet stream. Various buffer management schemes (such as RED [54, 10]) can also be employed on these tunnel queues to reduce the required buffer size and provide fair buffer allocation.

### 2.2.5.2  Control TCP Sender

The control TCP sender is the sender of a TCP connection set up between the sending and receiving nodes of a TCP circuit. In contrast with the normal usage, the control TCP sender is not a process running at the user-level. Instead, it refers to the TCP processing functions and the socket which represents the sending end point of the TCP connection in the kernel. The control TCP sender generates and transmits header packets, transmits data packets, and receive acknowledgment packets. All of these operations are automatically performed by the TCP processing functions, which are called by the network interrupt service routine, which in turn is invoked when a packet arrives. Since every operation is performed inside the kernel without context switching overhead between the kernel and user space, the control TCP sender operates efficiently and supports high speed forwarding.

To set up a control TCP connection between the sending and receiving nodes, like the normal usage, a user-level process at the sending and receiving nodes is run up. These two processes use the standard socket system calls such as connect() and accept() to conduct TCP's 3-way handshaking connection set up procedure. After the TCP connection is set up, the process on the sending node becomes idle and is not involved in sending header and data packets and receiving acknowledgment packets from the receiving node. Similarly, the process on the receiving node also becomes idle.

The socket send buffer [20] allocated to the control TCP sender is not used in the TCP decoupling approach as there is no physical data for the control TCP sender to send. Instead of working on and transporting a data byte stream formed by application data when they are written into a TCP socket send buffer as in the normal TCP usage, the control TCP sender works on and transports a "virtual data byte stream," which does not physically exist. Each packet transmitted by the control TCP sender thus is a packet consisting of only the TCP/IP header and contains no physical TCP data payload. They are thus called "header packets" in this thesis. These header packets, together with the acknowledgment packets sent back by the control TCP receiver, are called "control packets" as their existence are solely for congestion control purposes, rather than for data-carrying. The informa-tion carried in the TCP header of a header packet thus identifies some contiguous bytes of the virtual data byte stream that the header packet is supposed to carry, although physically they are not carried in the header packets.

The operations on the virtual byte stream closely correspond to the operations on the byte stream formed by the data packets entering the tunnel queue. When VMSS contiguous bytes of the virtual data byte stream has just been "transported" by the control TCP sender under its TCP congestion control, the corresponding VMSS bytes of the data packet stream can now be physically forwarded. Since data packets in the tunnel queue have many different sizes, forwarding VMSS bytes of the data packet stream actually translates to the forwarding of as many data packet as until these VMSS byte credits are exhausted. (Note that since each packet should be transmitted atomically and cannot be cut arbitrarily for transmission, sometimes credits may be left or overused by a little amount. These left credits or debits will be carried over to the next time when another VMSS bytes credits are gained again.)

One exception to the correspondence between the virtual byte stream and the byte stream formed by the data packets entering the tunnel queue is that, in case of a header packet loss, to keep TCP congestion control algorithms going, the control TCP sender must retransmit the lost "virtual" data until it is finally received by the receiving node (actually it is the lost header packet that matters). However, this retransmission operation does not result in a retransmission of the corresponding data of the data packet stream. Instead, taking advantage of these extra VMSS credits, more data packets totaling up to VMSS bytes are dequeued and forwarded out from the tunnel queue. This design is both desirable and simple. This design is desirable because no automatic retransmission of lost data packets is one important design goal of the TCP decoupling approach, which has been explained in Section 2.2.4. This design is simple because, now since data packets need not be retransmitted, the buffer space occupied by them can be released as soon as they are dequeued from the tunnel queue and forwarded out. There is no need to keep them in the tunnel queue as required in a TCP socket send buffer. This design therefore allows for a simple first-in-first-out buffer system for the tunnel queue.

For each header packet sent by the control TCP sender, the "CONTROL" bit is set in the Type-Of-Service (TOS) field of its IP header to allow the routers on the TCP circuit's path to distinguish header packets from data packets and thus be able to give them different treatments. Section 2.2.7 discusses the useful "lossless" property enabled by the use of this bit. In the TCP decoupling approach, setting the "CONTROL" bit does not need a complicated packet marker as required in the diff-serv approach [56], which needs to deliberately choose which packets to mark as "IN" or "OUT." The reason for this simple packet marking is that since all header packets are generated only by the control TCP sender and that the BSD kernel by default copies the TOS in a TCP connection's control block to the TCP connection's outgoing packets before they are sent, when a control TCP is set up, the standard setsockopt() system call can be used to turn on this bit and store it in the TOS field of the TCP control block of the control TCP in the kernel. From now on, the kernel automatically "marks" every header packet before sending it.

To meet the design goal of "data packets triggers control packets" as explained in Section 2.2.4, the generation and sending of header packets are enabled only when there are data packets to be forwarded in the tunnel queue and the control TCP's congestion control allows. In the TCP decoupling approach, an attempt to generate and send header packets is triggered every time a data packet arrives and is enqueued into the tunnel queue or the control TCP's congestion window size is opened up to allow more data packets to be sent by the arrival of an acknowledgment packet. However, only when the generation and sending of header packets are enabled, will a header packet be actually generated and sent out.

The sender of the control TCP uses the difference between its current congestion window size and the number of its current outstanding (not acknowledged yet) virtual bytes in the network as the credit to decide when it can forward more data packets and send them to the network. When the credit is below zero, no more data packets can be forwarded and sent to the network. Following the normal TCP design, when a control TCP is initially set up or when it times-out, the control TCP's congestion window size is set or reset to VMSS bytes. As a result, the control TCP sender always has VMSS bytes credits to transmit up to VMSS bytes data packets when it initially starts or restarts. For every VMSS bytes worth of data packets which have been forwarded and sent to the network, the control TCP sender sends out a header packet as if the header packet were coupled with these data packets as is performed in traditional TCP. The transmissions of these data packets precede the transmission of their associated header packet to meet the design goal of "data packets triggers control packets." The control information carried in the TCP headers of these header packets are exactly the same as the control information that would have been generated and carried if each header packet physically carries a VMSS-byte TCP data payload from a physical byte stream. The outcomes of these header packets, either successfully received and acknowledged or lost in the network, will cause the control TCP sender to adjust its congestion window size.

Multiple control TCPs can be set up between the sending and the receiving nodes of a TCP circuit to work together on the same data packet stream flowing into the TCP circuit. The senders of these control TCPs dequeue and forward packets from the same tunnel queue as soon as their TCP congestion controls allow them to send more data into the network. Using multiple control TCP connections is for two different purposes. First, using multiple control TCPs can smooth the achieved bandwidth usage of the TCP circuit (thus the data packet stream flowing in it). The reason is that if only one control TCP is used, since TCP reduces its sending rate at least by 50% when its packet gets lost and its fast retransmit gets triggered (explained in Section 2.1), the transmission rate of the data packet stream, which is regulated by the control TCP, will also undergo a 50% rate reduction. Suppose that there are now M control TCPs. Then a 50% of bandwidth reduction from any of them after TCP fast retransmit is triggered will only result in a reduction of the total bandwidth by a factor of (1/2)/M. This smoother bandwidth change is important for TCP trunking application where, in the backbone network, a trunk's achieved bandwidth should not vary too much and too quickly for stability concerns. Second, using multiple TCP connections is a way of allocating available bandwidth. It is well known that TCP exhibits per-flow fairness property [40, 55] -- i.e., when there are N greedy TCP flows with about the same round-trip time contending for available bandwidth, each one will roughly achieve 1/N of the available bandwidth. Using this property, a data packet stream regulated by N control TCPs can achieve N times bandwidth of a data packet stream regulated by only one control TCP.

The design of the architecture of the sending node of a TCP circuit has many useful properties. First, despite that multiple senders (one GMB sender plus one or multiple control TCP senders) can dequeue and forward data packets from the tunnel queue, the design maintains the packet order of the data packet stream when it flows through the sending node. That is, data packets in the tunnel queue are forwarded out in exactly the same order as they enter the tunnel queue. This in-sequence forwarding is an important property and is achieved by the design that when a control TCP sender decides to dequeue and forward a data packet from the tunnel queue, the control TCP sender must already have gained at least VMSS "credits" to forward a data packet. In case when a control TCP sender wants to dequeue a data packet and its current number of credits is less than the size of the first packet in the tunnel queue (this situation may happen when VMSS is configured to be smaller than links' MTUs), the control TCP sender simply returns and waits for more credits. No data packets will be queued in a control TCP sender as there is no need to queue data packets and no queue in a control TCP sender. A data packet thus can be sent to a network interface as soon as it is dequeued from the tunnel queue. Second, all operations (e.g., enqueueing data packets, dequeueing data packets, sending header packets, and receiving acknowledgment packets) are triggered and performed automatically in the kernel when packets arrive. This all-in-kernel design and implementation results in a high throughput system. Third, the format and content of data packets remain untouched and unchanged when they flow through the sending node.

### 2.2.5.3  GMB (Guaranteed Minimum Bandwidth) Sender

Consider the case when a data packet stream requires a guaranteed minimum bandwidth (GMB) of $X$ bytes per millisecond. Assume that via bandwidth provision and connection admission control, the network guarantees to deliver this required bandwidth for the data packet stream over its routing path. This section describes how a TCP circuit sends data packets at the GMB rate while being able to send additional data packets under TCP congestion control when extra bandwidth is available.

A TCP circuit has a GMB sender at the sending node of the TCP circuit. The GMB sender is equipped with a timer and unconditionally sends some number of data packets from the tunnel queue each time the timer expires. (In the current TCP decoupling implementation, the timer is set to be 1 millisecond.) When sending out data packets, the GMB sender need not send out header packets as a control TCP sender does. The reason is that the purpose of sending header packets is to probe for available bandwidth, and since the data packet stream has been allocated a certain bandwidth as its GMB, there is no need to send out header packets for data packets sent by the GMB sender. When the timer expires, if there are data packets in the tunnel queue, the GMB sender will send some of them under the control of a leaky bucket algorithm. The objective here is that, for any time interval of *Y* milliseconds, if there is a sufficient number of bytes to be sent from the tunnel queue, the total number of bytes actually sent by the GMB sender will approach the target of *X\*Y*.

For each expiration of the GMB timer, the GMB sender will try to send all the data packets it is supposed to send. If there are still some data packets left in the tunnel queue, they will be sent out under the congestion control of the control TCP sender(s) as described in Section 2.2.5.2. In this manner, the data packet stream will always receive its GMB under the control of the GMB sender, and at the same time dynamically share the available bandwidth under the congestion control of the control TCP(s).

Figure 5 depicts an ideal bandwidth allocation which is expected to result from using both one GMB sender and one control TCP sender for each of the three data packet streams in the simple network configuration. The achieved bandwidth of stream A should be 5 Mbps as it contains two parts. The first part is stream A's GMB of 2Mbps. The second part is stream A's fair share of the remaining bandwidth, which is (15 - 2 - 1 - 3)/3 = 3 Mbps. The same reason can be used for explaining the achieved bandwidths of stream B and C. Experimental results presented in Section 3.7 confirm this expectation.

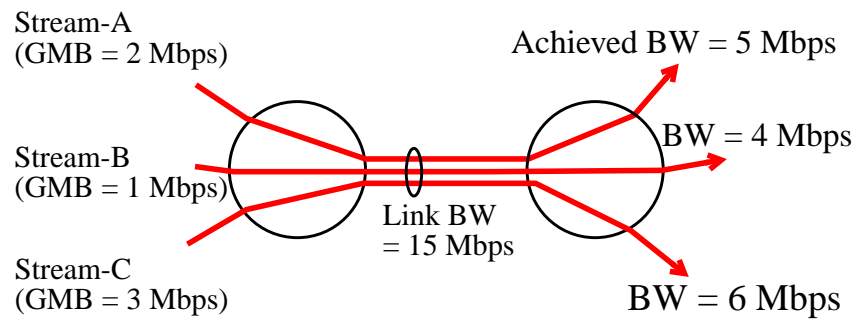

Figure 5. An ideal bandwidth allocation expected by using one control TCP sender and one GMB sender for stream A, B, and C in this simple network configuration.

## 2.2.6 The TCP Decoupling Mechanism on the Receiving Node of a TCP Circuit

Figure 6 depicts the architecture of the receiving node of a TCP circuit implemented in the TCP decoupling approach. The control TCP receiver is the receiver of a TCP connection set up between the sending and receiving nodes of a TCP circuit. In contrast with the normal usage, the control TCP receiver is not a process running at the user-level. Instead, it refers to the TCP processing functions and the socket which represents the receiving end point of the TCP connection in the kernel. The control TCP receiver receives header packets sent by its corresponding control TCP sender, and for each received header packet, the control TCP receiver views it as a TCP packet carrying VMSS-byte data payload, although physically there is no data payload coupled with the header packet. The control TCP receiver processes received header packets using the normal TCP cumulative acknowledgment scheme (explained in Section 2.1) and acknowledges their receipt by sending out acknowledgment packets. Since there is no real data payload carried in these received header packets, the control TCP receiver need not do a checksum test on the data payload, nor does it need to insert any data to its socket receive buffer. Receiving header packets and sending back acknowledgment packets are automatically performed by the TCP processing functions, which are called by the network interrupt service routine, which in turn is invoked when a packet arrives. Since every operation is performed inside the kernel without context switching overhead between the kernel and user space, the control TCP receiver operates efficiently and supports high speed forwarding.

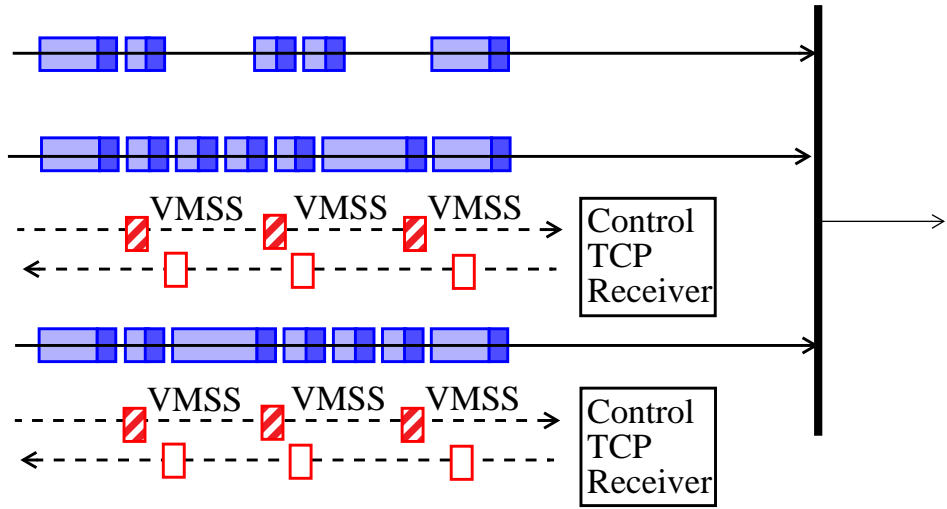

Figure 6. The architecture of the receiving node of a TCP circuit implemented in the TCP decoupling approach

To set up a control TCP connection between the sending and receiving nodes, like the normal usage, a user-level process at the receiving node is run up. This user-level process works with the user-level process at the sending node to conduct TCP's 3-way handshaking connection set up procedure. After the TCP connection is set up, the user-level process at the receiving node becomes idle and is not involved in receiving header and data packets and sending acknowledgment packets to the sending node.

Multiple control TCP receivers can be used, each corresponding to a control TCP sender at the sending node, to achieve the properties enabled by using multiple control TCP connections described in Section 2.2.5.2.

The design of the architecture of the receiving node of a TCP circuit has many useful properties. First, arriving data packets, either sent under the control of the GMB sender or the control TCP sender(s) at the sending node, are forwarded automatically by the kernel based on the IP destination addresses contained in their own TCP/IP headers. These data packets are forwarded in exactly the same way they would be forwarded in a normal router. No further processing on these data packets is needed. The control TCP receiver are not involved in the forwarding of these arriving data packets. (Actually the control TCP receiver does not even know when a data packet will arrive, nor does it know when a data packet has been forwarded out.) This design makes forwarding a data packet as fast as when the TCP decoupling approach is not used and results in a low-latency and high-throughput system. Second, the design maintains the packet order of a data packet stream when it flows through the receiving node. The reason is that, since data packets arrive in a sequential order and, as described above, each one can be forwarded out immediately, data packets thus will be forwarded out in exactly the same order as they arrive. Third, the content and format of data packets remain untouched and unchanged by the control TCP receiver when they flow through the receiving node.

## 2.2.7 Router Buffer Management Scheme for the TCP Decoupling Approach

In the TCP decoupling approach, the requirement for a router's buffer system can be as simple as a single FIFO queue shared by both data and header packets. A single FIFO queue allows for a simple and low-cost buffer system and preserves the order of arriving packets. To prevent loss of data packets in a FIFO queue, which is called the "lossless" property in the thesis, the router's buffer management system needs to ensure the following things:

- When the FIFO queue buildup occurs, drop some incoming header packets early enough so that their control TCP senders can reduce their rates of sending data packets in time.
- Allocate sufficient buffer space for data packets to accommodate temporary buffer usage fluctuation caused by the end-to-end control delay and possible arrival of new flows.

Figure 7 depicts the router buffer architecture using a single FIFO queue. It shows that, when there is no GMB traffic (the data packets sent under the control of GMB senders), the buffer space occupied by data packets is proportional (VMSS/HP_Sz times) to that occupied by header packets. Taking advantage of this property, controlling the maximum number of bytes of data packets in the FIFO can be achieved by limiting the maximum number of header packets in the FIFO. As a result, by properly controlling the maximum number of header packets so that the total buffer usage of the header and data packets is always below the provisioned buffer size, the TCP decoupling approach can achieve the "lossless" property for data packets.

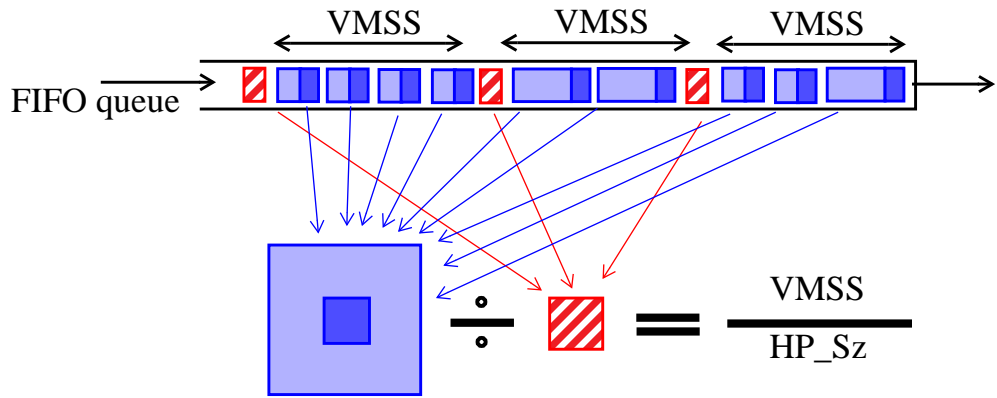

Figure 7. The router buffer architecture.

The maximum number of header packets in the FIFO is controlled by dropping them when the number exceeds a certain threshold. Since header packets are generated by control TCP(s), dropping header packets will trigger their control TCP senders' TCP congestion controls to reduce their sending rates. As a result, the buffer occupancy of header packets will drop below the threshold again and thus be maintained near the threshold. Section 2.2.7.1 will present the packet dropping method which controls the maximum number of header packets in the FIFO queue. Section 2.2.7.2 will present an analysis on how to provision adequate buffer space to ensure the "lossless" property when there is GMB traffic in the network.

### 2.2.7.1  Packet Dropping

The router uses the following policies to drop packets:
- Drop arriving header packets when their number in the buffer exceeds a certain threshold HP_Th.

- Attempt not to drop arriving data packets unless the buffer is really full. In fact, as will be explained in Section 2.2.7.2, with a size of Required_BS of Equation (2), the buffer can be guaranteed not to overflow.

Following the arguments of [48, 10], HP_Th is set to:

$$HP\_Th = \alpha * N \qquad (1)$$

where N is the number of active control TCPs that are expected to use the buffer at the same time, and $\alpha$ is the number of packets that the congestion window of a TCP connection must have in order to avoid frequent timeouts. A reasonable choice for $\alpha$ is 8 because if a TCP connection has 8 or more packets in its congestion window, chances that the fast retransmit and recovery mechanism [63] can recover from a packet loss due to TCP's ramp up to probe for available network bandwidth is pretty good. Because use of RED [54] can lower the value of $\alpha$ somewhat, in all experiments reported in this thesis, a simple RED-like scheme is used in the router for header packets.

Although both header and data packets may appear in the buffer, the criterion described above for dropping header packets depends only on the number of header packets in the buffer, and is independent of that of data packets in the buffer. Since this dropping mechanism is simpler than the dropping mechanisms used in diff-serv approaches [56], routers capable of supporting diff-serv approaches are expected to be able to support this packet dropping method as well.

### 2.2.7.2 Buffer Sizing

Given $\alpha$ and N, the next step is to compute the required buffer size, in bytes, to ensure no loss of data packets during congestion.

Let HP_Sz be the size of header packets in bytes. Recall that VMSS is the virtual maximum segment size in bytes. Typically, HP_Sz = 52 and VMSS = 1500.

Three types of packets may occupy the buffer of a router. This section discusses their buffer requirements in bytes separately:

1. *Header packets.*

    The required buffer size for these packets is:

    HP_BS = HP_Th*HP_Sz

2. *Data packets sent under the control of the control TCP(s)*

    The required buffer size for these packets is:

    UP_BS_TCP = HP_BS*(VMSS/HP_Sz) + N*VMSS

    The first term reflects the fact that a data packet is VMSS/HP_Sz times larger than a header packet. Because in the TCP decoupling design, for a control TCP sender, data packets totalling up to VMSS bytes are sent before their associated header packet is sent (explained in Section 2.2.5.2). The second term takes into account the worst case that every control TCP sender has forwarded out data packets totaling up to VMSS-1 bytes, but has not sent out the associated header packet because the number of bytes of data packets forwarded has not exceeded VMSS bytes.

3. *Data packets sent under the control of the GMB sender*

    Let the required buffer for these packets be UP_BS_GMB. Suppose that the fraction of the output link's bandwidth allocated for the GMB traffic is $\beta$, with $\beta < 1$. Since a FIFO queue can be viewed as an extension of the link and, at any time, $\beta$ of the link's capability which can hold packets being transmitted is occupied by the data packets sent under the control of the GMB sender, when the buffer is full, $\beta$ of the buffer space of the FIFO queue will be occupied by the data packets sent under the control of the GMB sender. That is,

    $\beta$ = UP_BS_GMB/(HP_BS + UP_BS_TCP + UP_BS GMB)

    Solving the above equation for UP_BS_GMB gives:

    UP_BS_GMB = (HP_BS + UP_BS_TCP)*$\beta$/(1 - $\beta$)

Thus the total required buffer size, Required_BS, to accommodate these three types of packets is:

**Required_BS**

$= HP\_BS + UP\_BS\_TCP + UP\_BS\_GMB$

$= (HP\_BS + UP\_BS\_TCP)*1/(1-\beta)$

$= (HP\_BS + HP\_BS*(VMSS/HP\_Sz) + N*VMSS)* 1/(1-\beta)$ (2)

where by Equation (1),

$HP\_BS = HP\_Th*HP\_Sz = \alpha*N*HP\_Sz$

Since a few percents drops of header packets are normally expected due to the control TCP's periodically probing for available bandwidth, the actual buffer requirement should be a few percents larger than Required_BS of Equation (2), to account for the fact that there could be a few percents more data packets than header packets in the buffer.

Like user packets sent by control TCPs, user packets sent by GMB senders need not be dropped in routers when congestion occurs. When deciding how much buffer to provide and reserve for GMB packets, Equation (2) already takes into account the maximum buffer space which GMB packets may occupy (UP_BS_GMB) when the provisioned buffer becomes full.

Thus, given the actual values for $\alpha$, $\beta$, N, HP_Sz and VMSS, Equation (2) estimates the buffer requirement that ensures no loss of data packets (either sent by control TCPs or GMB senders) during congestion. Experiments in Section 3.7 will demonstrate this lossless property. The small 5% difference between the estimated buffer requirement and the actual logged buffer usage in the experiments suggests the correctness and accuracy of the above buffer provision analysis.

## 2.3 Discussions about the TCP Decoupling Approach

### 2.3.1 Allocating Different Bandwidths to TCP Connections through the Use of Different VMSS Values

Control TCPs can use different VMSS values so that when they compete on a shared link, they can share the available bandwidth in proportion to their VMSS values. Due to the fact that in the TCP decoupling approach, when congestion occurs, routers will drop only header packets before dropping data packets, each competing control TCP will receive the same bandwidth for their header packets regardless of the number of bytes of data packet (VMSS) associated with each of their header packets. As depicted in Figure 8, suppose that the VMSS values of control TCP sender1 and control TCP sender2 are VMSS1 and VMSS2, respectively. Then the ratio of control TCP sender1's achieved bandwidth for its data packets to control TCP sender2's achieved bandwidth for its data packets will be VMSS1/VMSS2. Because of this property, using different values of VMSS for different control TCPs can allocate available bandwidths to these competing control TCPs in a fine-grain way. This property is particularly useful in TCP trunking application where it is desirable to allocate available bandwidth to different TCP trunks based on many different policies. For example, equal-sharing regardless of competing trunks' GMBs and proportional-sharing proportional to competing trunks' GMBs are just two special cases of what can be realized. The TCP trunking experiments suite TT1 (a) and (b) of Section 3.7 will illustrate this capability, where two values of 1500 and 3000 are used for VMSS.

When multiple control TCPs sharing the same buffer use different VMSS values, the value of VMSS used in Equation (2) for calculating the required buffer size should be the maximum of these VMSS values. Using the maximum of these VMSS values maintains the lossless property at the expense of increased required buffer space. If the physical buffer space is fixed and cannot be increased, two options are available. The first option is to use the maximum of these VMSS values but reduce the threshold HP_Th for header packets to still maintain the lossless property. Using a reduced HP_Th requires a reduction for either N or $\alpha$. That is, either a reduced number of control TCPs will use the buffer, or the control TCP sender's fast retransmit and recovery mechanism will work less well. The other option is to use the same HP_Th and allow data packets to be dropped sometimes when the buffer is full.

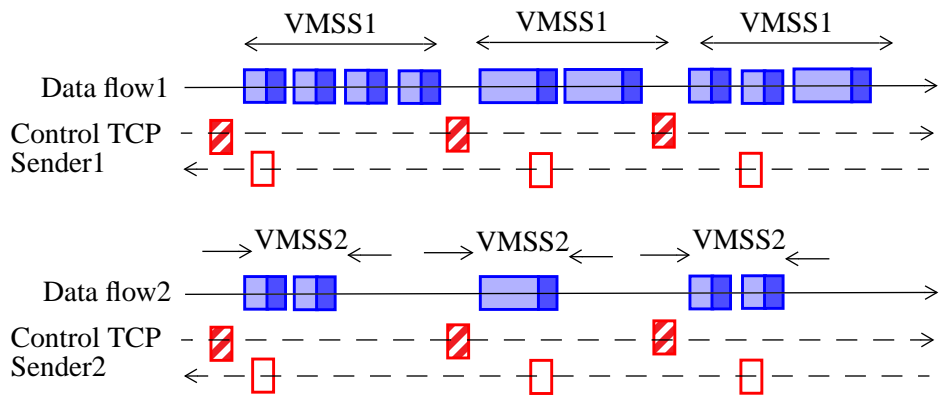

Figure 8. Although control TCP sender1 and sender2 achieve the same bandwidth for their header packets, the ratio of control TCP sender1's achieved bandwidth for its data packets to control TCP sender2's is VMSS1/VMSS2.

Yet another alternative is to use a RED-like scheme in the multi-VMSS values environment. The alternative is to use a RED-like buffer management scheme to replace the packet dropping and buffer sizing designed for the TCP decoupling approach. In the RED scheme, when the total buffer occupancy starts to exceed a certain threshold, incoming packets will be proactively dropped with a probability that is proportional to the current buffer occupancy level. That is, when the buffer occupancy increases, the probability that an incoming packet will be dropped also increases. This alternative slightly modifies RED so that only header packets are selected to be proactively dropped and data packets are forwarded without dropping unless the buffer space is totally consumed. Since dropping header packets can slow down control TCP sender's sending rates for data packets, which reduces the congestion and the current buffer occupancy level, the buffer can be maintained to not overflow, and thus data packets need not be dropped. This alternative is useful in the multi-VMSS values environment because when provisioning and configuring a router's buffer space, there is no need to know the VMSS values used by control TCPs in a network. Instead, for example, the proactive packet dropping threshold can be configured to be a half of the total buffer space and, as a result, the buffer will be non-empty at all times regardless of the VMSS values used. A buffer which is always non-empty means a high link utilization because there are always packets in the buffer to be sent to the link. This alternative achieves high link utilizations and ease of provisioning buffer space at the expense that the "lossless" property now is provided only with a high probability, but not 100%.

## 2.3.2 Accounting the Acknowledgment Packets Generated by Control TCP Receivers as Data Packets

Between two nodes in a network, there may be a control TCP in each direction. In this configuration, a control TCP sender and a control TCP receiver reside on the same node. For the control TCP receiver, after having generated an acknowledgment packet, it should send the acknowledgment packet to its corresponding control TCP sender immediately subject to no congestion control. Otherwise, slow returning acknowledgment packets will slow down the corresponding control TCP sender's potential sending speed. However, sending acknowledgment packets directly to the network without going through the tunnel queue of the control TCP sender residing on the same node in the reverse direction will overflow a router's buffer provisioned according to the analysis of Section 2.2.7.2. The reason is that sending every VMSS bytes of these acknowledgment packets is not associated with sending a header packet. As a result, the ratio of header packets and data packets (defined as non-header packets) in a router's buffer will exceed VMSS/HP_Sz and causes buffer overflow. The TCP decoupling design solves this problem by physically transmitting acknowledgment packets directly to the network interface but virtually transmitting them through the tunnel queue of the control TCP sender residing on the same node in the reverse direction. The detailed design is that an acknowledgment packet is accounted as a data packet of the control TCP sender in the reverse direction. When an acknowledgment packet is transmitted to the network interface, its packet length is "charged" to the control TCP sender in the reverse direction. Using this accounting, the control TCP sender in the reverse direction will correctly transmit a header packet each time after VMSS bytes of data or acknowledgment packets are transmitted.

### 2.3.3 Control Packet Overhead

Header packets, sent by the control TCP sender, are regarded as bandwidth overheads in the TCP decoupling approach because they do not carry data payloads and their existence is solely for congestion control purposes. The control TCP sender sends one header packet per VMSS-byte worth of data packets. Assume a typical situation where each header packet has HP_Sz = 52 bytes (40 bytes for the TCP/IP headers and 12 bytes for the TCP timestamp option) and VMSS is 1500 bytes (Ethernet's MTU). Then the header packet overhead ratio for data packets sent by the control TCP sender is HP_Sz/ VMSS = 52/1500, which is about 3.4%. In the reverse direction, the acknowledgment packets sent back by the control TCP receiver to the control TCP sender is also regarded as bandwidth overhead. Because in the TCP decoupling design (and also in the normal TCP design), a control TCP receiver sends back an acknowledgment packet for every other header packets, the acknowledgment packet overhead ratio is about (3.4% / 2), which is 1.7%. In total, the control packet overhead is 5.1%. The ratio can be lowered by increasing VMSS to a larger value.

VMSS can be larger than the path MTU without risking the possibility of packet fragmentation because the VMSS-byte worth of data packet(s) associated with a header packet is not sent out as a single IP packet of VMSS bytes. Instead, the data is sent as a sequence of separate data packets that are already queued in the tunnel queue. Traditionally it is unfavorable to use a large MSS (and thus a large MTU) to transfer a big chunk of packet in a network because, during its lengthy transmission, a packet with a higher priority such as voice cannot be transmitted. For example, since ATM [8] was designed with an aim to support real-time telephone traffic well, ATM decided to choose a small cell size of 53 bytes. In the TCP decoupling approach, using a large VMSS does not result in a poor support for delay-sensitive real-time packets. The reason is that, since VMSS bytes of data is sent as a sequence of separate data packets, not a single VMSS-byte large packet, a high-priority packet can cut in and be transmitted as soon as the ongoing transmission of a low-priority data packet is finished. In the TCP decoupling approach, the support for real-time packets is as good as that in the traditional approach.

For data packets sent under the control of the GMB sender, there will be no header packet overhead because in this case no header packets are sent. Thus, if some fraction of the bandwidth of a link is used by GMB traffic, the overall header overhead for the link is reduced accordingly. For example, when 50% of a link's bandwidth is reserved for GMB traffic, the header overhead is further reduced from 3.4% to 3.4% / 2 = 1.7%.

# Chapter 3  Application 1: TCP Trunking

## 3.1 Introduction

Trunking service has been used in networks for a long time to aggregate traffic with a common property and treat the traffic as a single entity. For example, all phone call traffic originating from Boston to New York can be bundled into a trunk starting in Boston and ending in New York, and the trunk traffic is routed as a whole to New York. The switches or routers on the path from Boston the New York can route the trunk traffic with a common destination to New York without looking at the phone number of each individual phone call. In New York, where the trunk ends, each individual phone call traffic then will be routed to its destination phone set based on its own phone number.

Aggregating network traffic with a common property into a trunk has many advantages. The first advantage is that trunking reduces the complexity of resource planning in networks. The unit of entities that need to be scheduled and planned now is a trunk rather than an individual traffic flow. The second advantage is that routing and switching become more efficient. Due to many other advantages which will be presented later, trunking service is being used extensively in networks.

This chapter introduces and presents TCP Trunking [24] and shows many of its offered advantages. A TCP trunk is a trunk whose traffic transmission rate is regulated by TCP congestion control. Section 3.2 discusses some related work. Section 3.3 gives an overview of TCP trunks. Section 3.4 presents the implementation of TCP trunks using the TCP decoupling approach. Section 3.5 presents many useful properties of TCP trunks. Section 3.6 discusses how to manage buffers at a TCP trunk sender. Section 3.7 presents many experiment results, and finally Section 3.8 considers the management and migration issues about TCP trunks.

## 3.2 Related Work

In ATM [8] and Frame Relay [21] transport networks, which are connection-oriented networks, the traffic of a flow must be carried on a virtual circuit (VC). A VC is established by configuring every switch on the path from the traffic source to the destination so that each switch knows its downstream switch when forwarding the traffic of this VC. The virtual path (VP) concept and mechanism is introduced to aggregate the traffic of multiple VCs using the same network path and the role of a VP is like a trunk. Associated with each VC and VP is an index called VCI and VPI, respectively. The VCI and VPI are contained in the header of each ATM or Frame Relay cell. The VCs that are aggregated in the same VP are given the same VPI of that VP and their traffic is routed based on the same VPI, rather than their own VCIs, on the VP path.

Recently, multi-protocol label switching (MPLS) approaches [51, 9] have been proposed in Internet to aggregate traffic using the same routing path so that IP route lookup operations can be avoided or facilitated. In this scheme, a label-switched path (LSP) is set up for the traffic traversing the same routing path and its role is like a trunk. When a packet enters a label-switch path, the first router on the LSP will perform a normal route lookup to determine the next hop for the packet. Also, a small label (functions like a VPI) is prepended to the packet before the packet is forwarded to the downstream router. The downstream router uses the label to quickly determine the next hop without performing a costly IP route lookup. Before forwarding the packet to the next downstream router, the downstream router replaces the label with a new one. This process repeats until the packet leaves the label-switched path and at that time the label is removed. This approach is similar to the routing approach based on the VCI/VPI used in ATM and Frame Relay networks.

In an ATM or Frame Relay network, the sending rate of an available-bit-rate (ABR) VP is adjusted dynamically to adapt to the current congestion in networks [1]. The goal is to achieve high link utilizations and low cell drop rates while allowing each VP to achieve its fair share of available bandwidth. Many rate-based [18] and credit-based [25] schemes have been proposed to try to achieve this goal. In the MPLS scheme, the sending rate of traffic traversing on a LSP is suggested to be regulated based on TCP's congestion control principles. Presently, however, no implementation has been proposed in the documents [51, 9] about how to use TCP congestion control to regulate the transmission rate of a LSP.

The TCP trunk approach proposed in the thesis is well suited to regulate the sending rate of traffic traversing on a LSP using TCP congestion control. The reason is that TCP trunks use genuine and unmodified TCP congestion control algorithms to regulate their transmission rates. Given the fact that TCP's congestion control is more well studied and sophisticated than the ABR rate-based congestion control algorithms used in ATM networks for a VC or VP, it can be advantageous to replace the ABR rate-based congestion control algorithms with the TCP decoupling approach for a VC or VP.

## 3.3 Overview of TCP Trunks

A TCP trunk is a layer-2 ATM or Frame Relay virtual circuit or an MPLS label switched path whose sending rate is under TCP congestion control. Optionally, via admission control and resource reservation, a TCP trunk can be allocated a GMB and, in this case, TCP congestion control is used to let the TCP trunk achieve available bandwidth beyond its GMB. Figure 9 (a) depicts an IP network with four router nodes. Figure 9 (b) shows two TCP trunks: tcp-trunk-1 from A to C and tcp-trunk-2 from D to B. The layer-2 circuit or MPLS label switched path associated with a TCP trunk is called the path of the trunk. For example, the path of tcp-trunk-1 is from A to B and to C. The sender and receiver of a TCP trunk are, respectively, the source and destination nodes of the trunk path. For example, tcp-trunk-1's sender and receiver are A and C, respectively. Like a conventional leased line or a layer-2 virtual circuit, a TCP trunk may carry a number of user flows, which are host-to-host TCP or UDP flows using the trunk path as parts of their routes. Packets of user flows are called user packets. Figure 9 (c) depicts that tcp-trunk-1 carries two user TCP flows, tcp-1 and tcp-2.

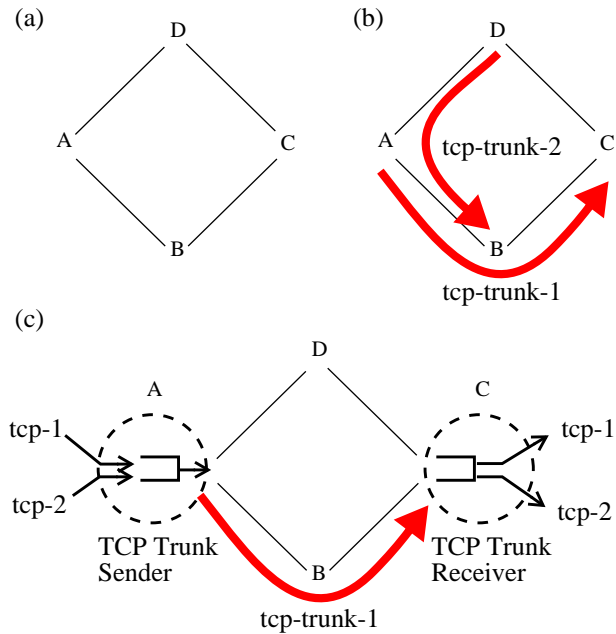

Figure 9. (a) An IP network; (b) two TCP trunks over the network; and (c) two user flows (tcp-1 and tcp-2) over tcp-trunk-1.

Setting up a TCP trunk over a layer-2 circuit or an MPLS path involves only configuring the two end nodes of the trunk. No states need to be maintained in the routers along a TCP trunk's path inside the network.

TCP trunking has many useful uses. For example, by aggregating a number of user TCP flows into a single TCP trunk, a TCP trunk can reduce the number of TCP flows a network router needs to handle, and thereby decrease packet drop rates [48]. By using a TCP trunk to carry UDP flows, which are not flow controlled and may not be TCP-friendly, UDP flows no longer can starve competing TCP flows. Since a TCP trunk is a layer-2 virtual circuit or an MPLS label switched path whose sending rate is under TCP congestion control, TCP trunks can dynamically adjust their bandwidth usages to keep the network utilization high and packet drop rates low.

## 3.4 Evolution of the Implementation of TCP Trunks

Several methods of implementing TCP trunks have been studied and experimented with PC machines running the FreeBSD 2.2.7 kernel. These methods can be characterized by how a TCP trunk carries user packets. Figure 10 depicts three of these methods. During the course of this research, the user-level method was first studied, then the encapsulation method, and finally, the decoupling method which, as this thesis argues, is the most attractive of the three methods.

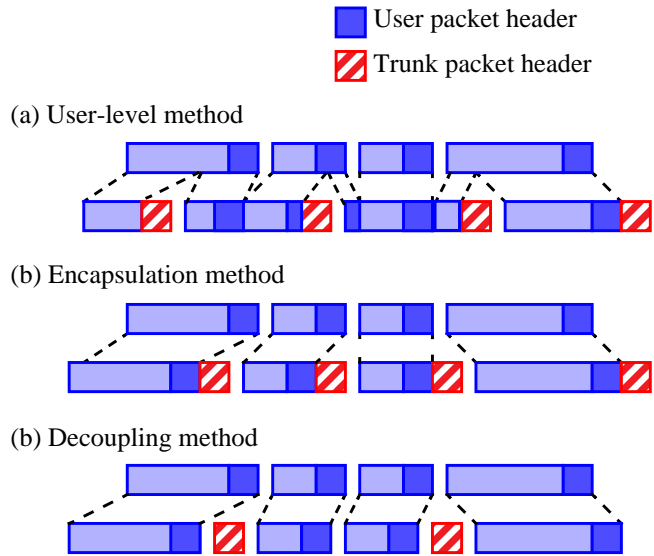

Figure 10. Three methods of implementing TCP trunks in terms of how a trunk carries user packets.

### 3.4.1 Early Implementation I: User-level Method

The user-level method is conceptually the most straightforward method of implementing TCP trunks. At the sender of a trunk, a user-level process, called "trunk application sender", will receive incoming user packets from the kernel, form a byte stream from these packets, and then write the byte stream to a TCP socket to transmit the byte stream over a TCP connection set up for the trunk. Packet boundaries between user packets in the byte stream are maintained by prepending a 2-byte length field to each user packet. As in any TCP connection, a TCP/IP header, called "trunk packet header" here, is prepended to each TCP segment sent over the trunk, as depicted in Figure 10 (a). At the trunk receiver, a user-level process, called "trunk application receiver", will receive these bytes from the TCP socket of the TCP connection of the trunk, recover user packets using their length information, and then send them to the kernel. The kernel will then forward these packets out to the proper network interfaces.

Although conceptually straightforward, this user-level method has two serious drawbacks. First, the trunk application sender and receiver suffer from the overhead of copying every user packet up into the user space and then copying it down to the kernel space. Second, TCP's insistence on offering a reliable and in-sequence delivery service causes forwarding delays at the trunk receiver when a user packet is lost. That is, the TCP processing module in the kernel at the trunk receiver cannot deliver a received user packet from the kernel to the user space for the trunk application receiver until any lost byte that the trunk application sender transmitted earlier has been successfully recovered via TCP retransmission. Thus a received user packet may be stuck in the kernel for a long period before being delivered to the user-level trunk application receiver to be forwarded out to the next hop.

### 3.4.2 Early Implementation II: Encapsulation Method

Under the encapsulation method, the TCP trunk sender directly enqueues incoming user packets into the TCP's socket send buffer in the kernel space when they arrive. This design avoids the two data copy operations required in the user-level implementation of Section 3.4.1. The TCP processing module at the trunk sender is modified to take exactly one complete user packet as the payload of each outgoing trunk packet. In other words, each incoming user packet is encapsulated in a trunk packet using the TCP/IP header associated with the TCP connection of the trunk. Figure 10 (b) illustrates the encapsulation.

Since each received trunk packet now contains one complete user packet, the trunk receiver can easily extract the user packet from the trunk packet, make a copy of it, and forward the copy out immediately, without waiting for any other packet. The normal TCP processing module will process each received trunk packet (e.g., generating a duplicate acknowledgment packet when there are packet losses) as if the copy and forwarding operations never took place. Thus, there is no need to modify the normal TCP processing module at the trunk receiver. Those user packets which are later delivered to the socket receive buffer from the TCP processing module's assembly queue will be simply discarded, as they have already been copied and their copies have been forwarded out.

Under the encapsulation method, after a trunk packet is lost and before its repair is received, the trunk receiver can still continuously forward other received user packets (their copies) as they arrive. Although at the receiving host of user packets, user packets may arrive out of order due to the TCP trunk's retransmission of lost trunk packets, this continuous forwarding feature can be useful for those applications, such as video streaming, which demand predictable low-delay packet delivery and can tolerate a modest amount of out-of-order packet delivery.

The encapsulation method, however, has some shortcomings. First, since each trunk packet contains only one user packet, the trunk packet header overhead is significant for small user packets. Second, adding the trunk packet header to a large user packet may result in a packet of a size exceeding the MTUs of the links on the trunk path. An oversized packet will cause packet fragmentation and thereby introduce processing overheads and delays at the trunk sender and receiver. In addition, network bandwidth usage will become inefficient as in an IP network, if any fragment of a packet is lost, the whole packet needs to be retransmitted. Third, requiring TCP to recognize packet boundaries when sending data violates the semantic of TCP as a byte-stream protocol and makes the trunk sender's implementation unnatural and complicated. Finally, because a TCP connection has a non-zero inherent packet loss rate for periodically probing for available bandwidth (explained in Section 2.1), some user packets encapsulated in trunk packets need be dropped and retransmitted. However, this kind of inherent packet dropping can be avoided in the TCP decoupling method described in the next section.

### 3.4.3 The Chosen Implementation: Decoupling Method

A TCP circuit implemented in the TCP decoupling approach can be readily used as a TCP trunk when the TCP circuit's routing path is fixed either via a layer-2 virtual circuit or an MPLS label-switched path. The sender and receiver of a TCP trunk corresponds to the sender and receiver of a TCP circuit, respectively. The user packets transmitted over a TCP trunk correspond to the data packets transmitted over a TCP circuit. Actually, a TCP circuit was named "TCP trunk" when the TCP decoupling approach was originally developed solely for the TCP trunking application. Later on, when it was found that a TCP circuit is useful for a broad range of applications, not just for the TCP trunking application, the name of "TCP trunk" was changed to "TCP circuit" to make the name not application-specific.

The TCP decoupling method removes the drawbacks of the user-level and encapsulation methods discussed earlier in Section 3.4.1 and Section 3.4.2. First, when user packets arrive at the TCP trunk receiver node, they can be independently and immediately forwarded out. There will be no packet forwarding delay at the trunk receiver caused by TCP's insistence on providing a reliable and in-sequence delivery service. Second, the trunk sender inserts a header packet after having transmitted VMSS bytes of user packets. Thus, by choosing a sufficiently large VMSS value, the overhead of header headers can be made to be a fixed low value, independent of the distribution of user packet sizes. Third, in contrast to the user-level and encapsulation methods, the TCP decoupling method does not add any extra bytes to a user packet. Therefore, it does not cause packet fragmentation due to increased packet size. Fourth, the TCP decoupling method does not require TCP to recognize user packet boundaries when sending data, and thus leads to a simple implementation. Finally, assuming routers on the trunk path can provide the buffer management scheme designed for the TCP decoupling approach presented in Section 2.2.7, user packets need not be dropped due to congestion or TCP ramp-up.

## 3.5 Properties of TCP Trunks

A TCP trunk implemented in the TCP decoupling approach can simultaneously satisfy a number of properties of interest to various applications. These include:

- *Guaranteed and elastic bandwidth*
  - Guaranteed minimum bandwidth (GMB): A TCP trunk can guarantee that it will deliver at least some number of bytes of data over a period of a time when there are data to be sent.
  - Elastic bandwidth: Beyond GMB, a TCP trunk can use additional network bandwidth when it is available. A TCP trunk can share the available bandwidth with other competing TCP trunks in a fair way, in proportion to the trunk's GMB, or in any other desired proportion.

- *Immediate and in-sequence forwarding*
  - At the trunk sender, arriving user packets will be immediately forwarded to the trunk, unless they are flow controlled by the trunk's TCP congestion control in response to the congestion on the trunk path. Similarly, at the trunk receiver, arriving user packets will be immediately forwarded to output network interfaces. That is, they will be forwarded immediately without waiting for any other user packet which may be delayed or lost.
  - User packets arriving at the trunk sender or receiver will be forwarded out in-sequence, that is, in the order of their arrivals.

- *Lossless delivery*
  - Suppose that routers on the path of the trunk can differentiate trunk and user packets by their packet marking, and during congestion, these routers can drop trunk packets rather than user packets. (This mechanism is similar to what routers supporting diff-serv [56] can do.) Then the TCP trunk can guarantee that user packets carried by the trunk will not be dropped due to buffer overflow in these routers, while being able to adapt its bandwidth of transmitting user packets to network congestion level using TCP's congestion control. (Note, however, that user packets may still be dropped at the trunk sender if they arrive at a rate higher than the achieved bandwidth of the trunk. Section 3.6 discusses the buffer management at the trunk sender.)

- *Aggregation and isolation*
    - By aggregating a number of user flows into a TCP trunk, the number of flows which the routers on the trunk path needs to handle is reduced. As a result, packet drop rates on these routers are also reduced [48].
    - By using a TCP trunk to carry UDP flows, which are not flow controlled and may not be TCP-friendly, these UDP flows no longer can starve competing TCP connections.
    - By aggregating TCP flows from various user sites in separate TCP trunks, sites with different numbers of flows can share the network bandwidth fairly.
- *Easy set up and configuration*
    - Setting up a TCP trunk involves only configuring the two end nodes of the trunk.

## 3.6 Trunk Sender Buffer Management

The sender of a TCP trunk needs to buffer user packets if they arrive at a rate higher than the achieved bandwidth of the trunk. When the buffer is full, arriving user packets will have to be dropped. This phenomenon is similar to the case that the sender of a fixed-bandwidth leased line needs to buffer or drop data packets when they arrive at a rate higher than the bandwidth of the leased line. The TCP trunk's situation is more subtle than the leased line's situation because the achieved bandwidth of the TCP trunk is subject to the congestion control of its control TCPs and thus may vary over time.

Since TCP trunks are designed to carry aggregate user flows in a backbone network, to provide a stable quality of service (QoS) to end user applications, a TCP trunk's achieved bandwidth should not decrease too much and too quickly. This design goal can be achieved by using multiple control TCPs for a TCP trunk as explained in Section 2.2.5.2. In the situation when some arriving user packets need to be dropped because their aggregate arriving rate is greater than the TCP trunk's achieved bandwidth, fairness should be considered when selecting which user packets to drop unless some special QoS scheme (e.g., per-flow queueing and weighted round-robin scheduling) is implemented.

This section assumes that all user flows of a TCP trunk are TCP flows, a single FIFO is used as the tunnel queue at the TCP trunk sender, and a single control TCP is used for a TCP trunk. This section will not discuss the case that the sender of a TCP trunk uses a single FIFO as its tunnel queue and the TCP trunk's user flows consist of both TCP and UDP user flows. In such a case, because per-flow queueing and weighted round-robin scheduling are not used at the TCP trunk sender, non-flow controlled UDP user flows of the TCP trunk will starve user TCP flows of the same TCP trunk and use all of the TCP trunk's achieved bandwidth. Since this unfair bandwidth sharing problem can be solved by separating user UDP flows from user TCP flows and using one TCP trunk to transport user UDP flows and another TCP trunk to transport user TCP flows (see Experiment TT3 (b) of Section 3.7.3), this session will not discuss the mixed TCP and UDP user flows case. In the configuration that all user flows of a TCP trunk are all TCP flows, a single FIFO is used as the tunnel queue at the TCP trunk sender, and a single control TCP is used for a TCP trunk, this session focuses on the interaction of the two levels (i.e., the trunk and user flow levels) of TCP congestion control, and describes a solution for fairly allocating a TCP trunk's achieved bandwidth among its user TCP flows.

Consider the situation when a header packet of the control TCP of the trunk is dropped on the trunk path due to congestion. After recognizing this packet loss, the trunk sender will reduce the trunk's sending rate of user packets by 50% due to the triggering of TCP fast retransmit and recovery mechanism. In the meantime, a user flow on the trunk may not necessarily experience any packet loss and thus may continue transmitting at the same or even increased rate (explained in Section 2.1), in spite of the fact that the underlying trunk has already shrunk its sending rate. The number of the user flow's packets in the trunk sender's FIFO queue will therefore increase and the user flow's packets will eventually be dropped when the FIFO queue overflows. At that time, this user packet dropping will trigger the TCP congestion control at the sender of the user flow and reduce the sending rate of the user flow by at least 50%.

Ideally, when the TCP trunk sender reduces its bandwidth by some factor, all of the active user TCP flows over the trunk should also reduce their bandwidths by the same factor. The following three principles are used to provide an approximate solution for achieving this objective:

P1. The trunk sender needs to have a buffer of size about RTTup*TrunkBW/2, where RTTup is an upper estimate of RTTs of user TCP flows and TrunkBW is the target peak bandwidth for the TCP trunk, to store user packets when the TCP trunk suddenly reduces its sending rate by 50% due to fast retransmit and recovery mechanism. The reason is that, at any time, the TCP trunk has at most RTTup*TrunkBW outstanding bytes in the network. When the number of in-flight user packets on the TCP trunk path is reduced from RTTup*TrunkBW to RTTup*TrunkBW/2 due to the 50% rate reduction, the number of user packets which need to be queued at the TCP trunk sender is at most RTTup*TrunkBW - RTTup*TrunkBW/2 = RTTup*TrunkBW/2. Therefore, a buffer of size RTTup*TrunkBW/2 is large enough to hold user packets without excessive packet droppings when the TCP trunk reduces its sending rate by one half.

P2. Most of the time, the buffer occupancy of user packets at the TCP trunk sender should be maintained below a threshold, which is a low value. A RED-like packet dropping policy can be used to proactively drop user packets when the buffer occupancy exceeds this low threshold. Maintaining the buffer occupancy at a low level reserves buffer space to absorb a sudden high demand for buffer space when the TCP trunk suddenly reduces its sending rate by 50% and RTTup*TrunkBW/2 bytes of user packets need to be stored in the buffer. When the buffer usage passes the threshold due to the sudden trunk bandwidth reduction, the trunk sender drops some user packets to signal congestion to their TCP senders.

P3. After the trunk sender has dropped a packet from a user flow, the trunk sender will try not to drop another packet from the same user flow, until the user flow has recovered from this packet loss by fast retransmit and recovery mechanism. Note that for a user flow, the TCP fast retransmit triggered by the dropping of one of its packets will cause the user flow to reduce its transmission rate by one half. This rate reduction matches that of the underlying trunk when the TCP trunk's control TCP probes for available network bandwidth, loses a single header packet, and the trunk sender's TCP fast retransmit gets triggered. The user flow's 50% rate reduction is even too much when the TCP trunk is implemented using more than one control TCPs because in that case the trunk's bandwidth reduction is less than 50% (explained in Section 2.2.5.2). Additional packet drops from the same user flow are likely to only cause unnecessary TCP time-outs for the user flow. Therefore, the goal is that, every time the underlying TCP trunk shrinks its bandwidth under TCP congestion control, every active user TCP flow of the trunk should be signaled to reduce its sending rate by dropping one of its packets, and the dropping policy should allow TCP fast retransmit and recovery mechanism to work for all active user TCP flows whose packets are selected to be dropped.

To implement the P3 principle, a simple per-flow packet accounting method is used. The trunk sender estimates the total number U of packets that can be sent by a user TCP flow sender between the time right after the user TCP flow sender reduces its sending rate by one half and the time its sending rate is about to ramp up to its previous sending rate before its packet was dropped. This number U is used to set a threshold K, which will be the minimum number of packets from the user TCP flow that should be forwarded without being dropped, before any packet from the same flow will get dropped again.

Suppose that the user TCP flow's congestion window has w packets just before its fast retransmit is triggered. The number U of packets sent between the time right after the user TCP flow sender reduces its sending rate by one half and the time its sending rate is about to ramp up to its previous sending rate before its packet was dropped is roughly w/2 + (w/2+1) + (w/2+2) + .... + w = (3/8)*w^2. Since dropping another packet from this user flow too soon may fail the user TCP flow's fast retransmit and recovery mechanism, the threshold K should not be too small. On the other hand, if K is set to be too large, a user flow's exemption period, in which the user flow can keep growing its sending rate despite the current network congestion, could be too long. This long exemption period will make congestion control less responsive. For the experiments reported in this thesis, the value of K is set to be U/2. Measured performance results were found to be not sensitive to the precise value of K. That is, any value close to U/2 can be used as K to achieve similar results.

When a TCP trunk shrinks its bandwidth, the sender of the TCP trunk determines the value of K, which will be used for every active user TCP flow, as follows. Let W be the congestion window size of the control TCP of the TCP trunk right before its fast retransmit is triggered. By tracking the number N of active user TCP flows on the TCP trunk, the congestion window size of each active user TCP flow w when their packets are going to be dropped by the trunk sender due to the TCP trunk's bandwidth reduction is estimated to be W/N. Therefore, substituting W/N for w in the above formula, the threshold K is set to be (3/8)*(W/N)^2*(1/2).

## 3.7 TCP Trunk Experiments and Performance Measurements

Various TCP trunking experiments have been performed on the laboratory testbed networks. Some testbeds involve as many as 16 hosts and routers. The hosts and routers in the testbeds are FreeBSD 2.2.7 systems running on 300 MHz PCs each with 96MB of RAM and several Intel EtherExpress 10/100 cards set at 10 Mbps. A delay box imple-mented in the kernel is used to emulate a link's propagation delay. The RTT of a connection can be flexibly set to any desired value with a 1-ms granularity using the delay box.

Experimental results have validated the properties of TCP trunking listed in Section 3.5 such as providing elastic and guaranteed bandwidths, lossless data transport, and isolating UDP flows. Extensive simulation results generated by the Harvard TCP/IP network simulator [58] under various network configurations also confirm these properties. This section presents results from three suites of experiments.

### 3.7.1 Experiments Suite TT1: Basic Capabilities of TCP Trunks

This experiment suite demonstrates the basic capabilities of TCP trunks in bandwidth management as described in Section 2.3.1. The network of Figure 11 is used for this experiments suite.

Below are the configurations common to experiments TT1 (a), (b) and (c):

- Each trunk uses 4 control TCPs.
- Each trunk's input queue (tunnel queue) has a buffer size of 100 packets.
- The buffer in the bottleneck router E is of size Required_BS given by Equation (2) of Section 2.2.7.
- The bottleneck router E uses the packet dropping scheme presented in Section 2.2.7 to drop header packets when congestion occurs.
- The user flows are greedy UDP flows using 1,500 byte packets. (The case that 4 greedy TCP flows using 1,500 byte packets per TCP trunk was also tried and the results were similar.)
- The propagation delay of the link between E and F is 10 ms. The propagation delay of any other link is negligible.
- Each experimental run lasts 400 seconds or longer.

**Experiment TT1 (a):**

Configurations:

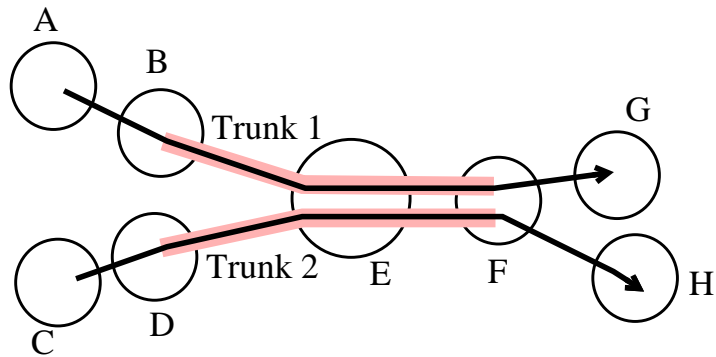

Figure 11. An experimental network testbed with 6 hosts and 2 routers, for TCP Trunking Experiments Suite TT1. Node E is the bottleneck router. The sender and receiver of Trunk 1 are B and F, respectively. The sender and receiver of Trunk 2 are D and F, respectively. Nodes A and G are the sender and receiver of a user flow using Trunk 1. Finally, nodes C and H are the sender and receiver of another user flow using Trunk 2. All links are 10 Mbps. Trunks 1 and 2 share the same 10 Mbps link from E to F.

- Trunk 1: GMB = 400 KB/sec, VMSS = 3000 bytes
- Trunk 2: GMB = 200 KB/sec, VMSS = 1500 bytes

An objective of this experiment is to demonstrate that, with the TCP decoupling approach, trunks can fully utilize available bandwidth and share it in proportion to their guaranteed minimum bandwidths (GMBs). This is achieved by choosing Trunk 1's VMSS to be twice as large as Trunk 2's VMSS, since Trunk 1's GMB is twice as large as Trunk 2's. The bandwidth allocation according to analysis should be:

Trunk1:

400 + 2/3 * (1200 - 400 - 200) = 800 KB/sec

Trunk2:

200 + 1/3 * (1200 - 400 - 200) = 400 KB/sec

For each of the above two equations, the first term is the trunk's GMB, and the second term is the extra bandwidth that this trunk should achieve when competing for available bandwidth with the other trunk. The available bandwidth is the remaining bandwidth on the bottleneck link (the link from E to F) after deducting Trunk 1 and Trunk 2's

GMBs (400 and 200 KB/sec) from the bottleneck link's bandwidth (10 Mbps = 1200 KB/sec). Since Trunk 1's VMSS is twice as large as Trunk 2's, Trunk 1 should achieve two times Trunk 2's bandwidth in sharing the available bandwidth. That is, Trunk 1 should achieve 2/3 of the available bandwidth and Trunk 2 should achieve 1/3 of the available bandwidth.

The experimental results, as depicted in Figure 12, show that each trunk achieves what the analysis predicts. That is, Trunk 1 and Trunk 2 achieve 800 and 400 KB/sec, respectively.

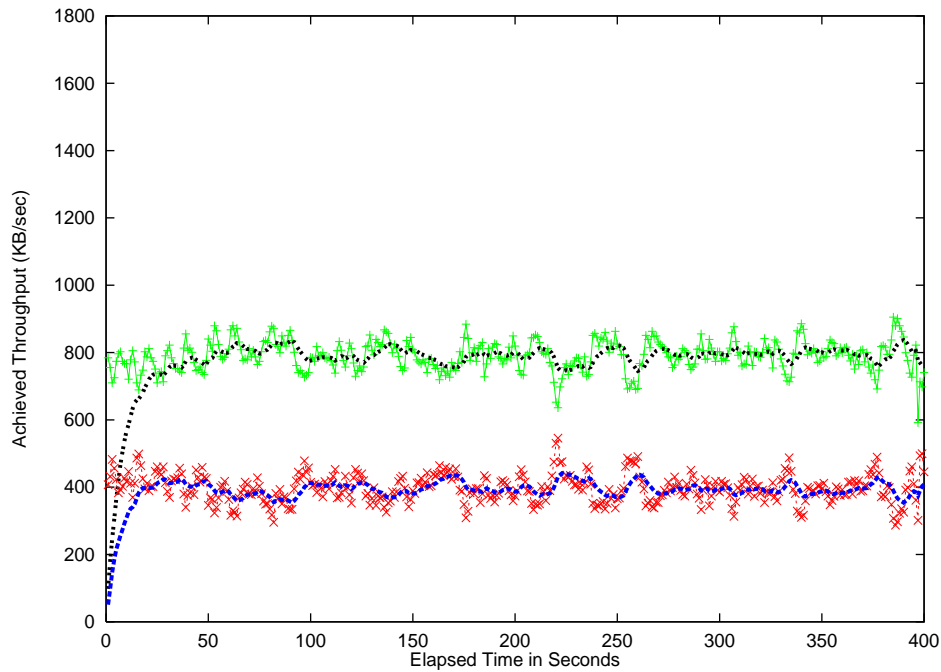

Figure 12. Results of Experiment TT1 (a). Each small point represents a trunk's achieved bandwidth averaged in that 1-second period around the point. The thick line represents the exponential running average of a trunk's achieved bandwidth over time. The achieved bandwidth of each trunk is exactly what the analysis predicts.

**Experiment TT1 (b):**

Configurations:

- Trunk 1: GMB = 200 KB/sec, VMSS = 3000 bytes

- Trunk 2: GMB = 400 KB/sec, VMSS = 1500 bytes

An objective of this experiment is to demonstrate that, with the TCP decoupling approach, trunks can fully utilize available bandwidth and share it in proportions which are independent of the trunks' GMBs. In this configuration, Trunk 1 has a larger VMSS value than Trunk 2, although the former has a smaller GMB than the latter.

Based on the same reasoning as that used in TT1 (a), the bandwidth allocation according to the analysis should be:

Trunk1:
   200 + 2/3 * (1200 - 400 - 200) = 600 KB/sec

Trunk2:
   400 + 1/3 * (1200 - 400 - 200) = 600 KB/sec

Again, the experimental results, as depicted in Figure 13, show that each trunk achieves about 600 KB/sec. This is what the above analysis predicts.

**Experiment TT1 (c):**

Configurations:

- Trunk 1: VMSS = 1500 bytes, GMB = 400 KB/sec
- Trunk 2: VMSS = 1500 bytes, GMB = 200 KB/sec

This experiment focuses on the buffer occupancy in the bottleneck router E and compares it with the Required_BS value given by Equation (2) of Section 2.2.7. The purpose of this experiment is to verify that there is indeed no loss of user packets in router E.

Using the notations of Section 2.2.7, the values of ($\alpha$, $\beta$, N, VMSS) used for this configuration is (8, 0.5, 8, 1500). The value of $\alpha$ is set to be 8 so that each control TCP's fast retransmit and recovery mechanism can work reasonably well. The value of $\beta$ is 0.5 because the sum of Trunk 1 and Trunk 2's GMB (400 + 200 = 600 KB/sec) is 50% of the bottleneck link's bandwidth (1200 KB/sec). N is 8 because Trunk 1 and Trunk 2 together have 8 control TCPs using the bottleneck router E's buffer. When plugging these values into Equation (2) of Section 2.2.7, the calculated Required_BS is 222,348 bytes.

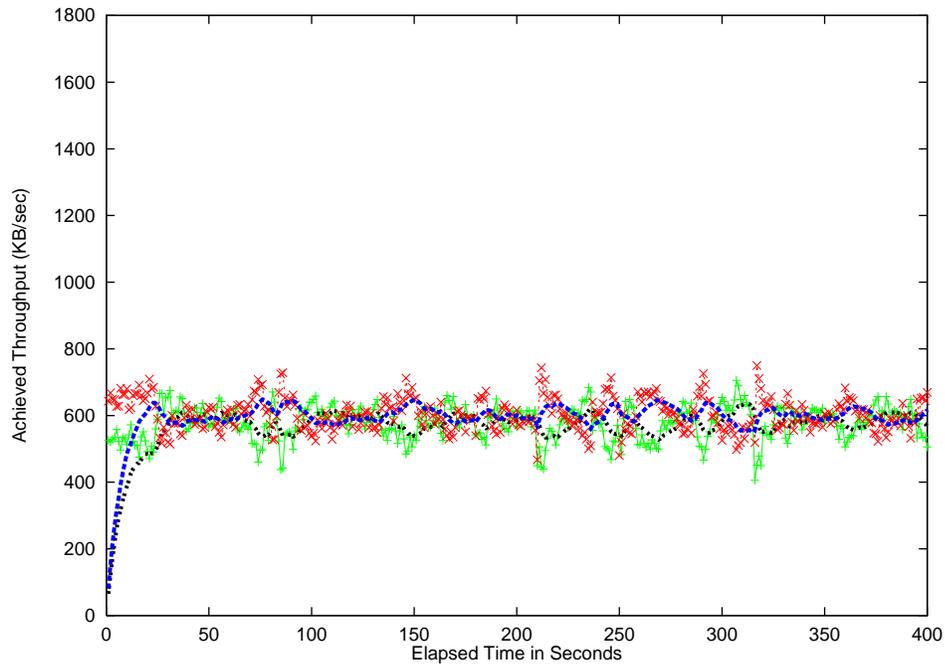

Figure 13. Results of Experiment TT1 (b). The achieved bandwidth of each trunk is exactly what the analysis predicts.

In the 600-second run, the logged maximum buffer occupancy is 210,306 bytes. Since the buffer of Required_BS or 222,348 bytes provisioned in the experiment is greater than 210,306 bytes, there is no loss of user packets. The fact that Required_BS of 222,348 bytes is only about 5% off from the maximum buffer occupancy of 210,306 bytes suggests the high accuracy of the calculation of Required_BS by Equation (2). Figure 14 depicts sampled buffer occupancy in the bottleneck router E during this experiment.

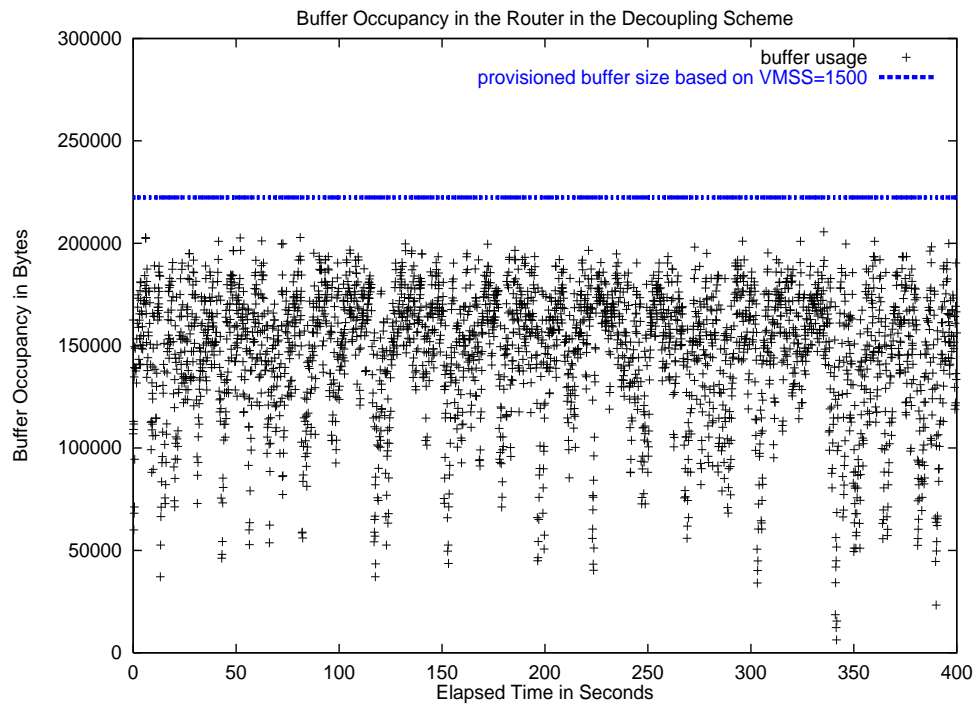

Figure 14. Results of Experiment TT1 (c). Sampled buffer occupancy in bytes in the bottleneck router E is shown. The top thick line is the Required_BS value given by Equation (2), i.e., 222,248 bytes. Note that sampled buffer occupancy is always below the line. In fact, the logged maximum occupancy is 210,306 bytes. Thus, in the experiment there is no loss of user packets.

In summary, the results of experiments TT1 (a), (b) and (c) show that a TCP trunk can:

- Guarantee GMB.

- Use multiple control TCPs to smooth bandwidth adaptation. Some other experiments not shown here have demonstrated that the curves in Figures 12 and 13 exhibit much larger degrees of variations if each TCP trunk sender uses only one control TCP.

- Use various values of VMSS to achieve fine-grain bandwidth allocation. Equal sharing and proportional sharing based on trunks' GMBs are just two special cases of many that can be achieved.

- Provide lossless delivery. As an evidence of the lossless property, the experiments have shown that the maximum buffer occupancy in bytes in the bottleneck router E is always bounded above by the Required_BS value given by Equation (2) of Section 2.2.7. This result is depicted in Figure 14.

### 3.7.2 Experiments Suite TT2: Protection for Interactive Web Users

This suite of experimental results, depicted in Figure 15, shows that TCP trunking can provide protection for interactive Web users when they compete against long-lived greedy TCP connections, i.e., short Web transfers can receive approximately their fair share of the available bandwidth and avoid unnecessary timeouts. In these experiments, each run lasts 10 minutes or longer.

Consider the configuration depicted in Figure 15 (b). On the middle router where traffic merges, there are many short-lived web transfers coming from an input port (a site) to compete for an output port's bandwidth (1100 KB/sec) with other long-lived greedy ftp transfers that come from two other input ports (sites).

Figure 15 (a) shows that when there are only short-lived 8KB web transfers in the network, the offered load uses 453 KB/sec bandwidth. (The offered load is limited to 453 KB/sec, because TCP windows for these web transfers never grow up significantly, due to the small 8KB size of the transfers.) The request-response delays for these short-lived web transfers are small and predictable. The mean delay, maximum delay, and the standard deviation of the delays are 353 ms, 1,270 ms, and 82 ms, respectively.

Figure 15 (b) shows that after long-lived greedy ftp transfers ("put file" sessions) are introduced into the network, the short-lived web transfers can only achieve 122 KB/sec bandwidth in aggregate, which is much smaller than its fare share (1200/3 KB/sec). The mean delay, maximum delay, and the standard deviation of the delays increase greatly and become 1,170 ms, 11,170 ms, and 1,161 ms, respectively. These worsened performance metrics indicate that the short-lived web transfers are very fragile (the reasons are discussed in [11]) and encounter more time-outs than before. As a result, the short-lived web transfers cannot receive their fair share of the bandwidth of the bottleneck link when competing with long-lived greedy ftp transfers.

Figure 15 (c) shows that when a TCP trunk is used for each site to carry the site's aggregate traffic, the bandwidth used by the short-lived web transfers increases to 238 KB/sec. The mean delay, maximum delay, and the standard deviation of the delays also improve greatly and become 613 ms, 2,779 ms, and 274 ms, respectively.

### 3.7.3 Experiments Suite TT3: Trunk Performance on a Ring

This experiments suite measures performance of TCP trunks on a ring. The experiments show that TCP trunking provides the following functions. First, it protects small TCP transfers against large ones. Second, it protects TCP flows against UDP flows. Third, it protects sites with a small number of TCP connections against those sites with a large number of TCP connections. This experiment suite uses a ring testbed network of Figure 16, which can test the performance of TCP trunks under multiple bottlenecks.

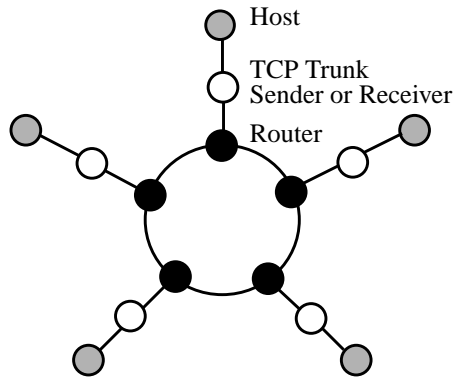

Figure 16. A ring testbed network for TCP trunking experiments TT3. The testbed consists of five hosts, five edge routers which are used as TCP trunk senders or receivers, and five routers on the ring.

As depicted in the figure, the testbed has five routers on the ring, five edge routers where the senders and receivers of TCP trunks are implemented, and five hosts where senders or receivers of user TCP or UDP flows reside.

All the experimental runs last 300 seconds. Each of these routers is configured to have a buffer of 50 packets for header packets, and each trunk sender a buffer of 100 packets for user packets. All the links on the testbed have negligibly small propagation delays. The maximum window size for user TCP flows is 64KB.

**Experiment TT3 (a): Use of Trunks to Protect Small TCP Transfers**

In this experiment, as depicted in Figure 17, there are ten ftp clients at node 1 that generate "get" requests to a ftp server at node 2. These "get" requests simulate web requests for small files. To reduce the effect of synchronization, ftp clients request files of different sizes of 8KB, 16KB and 32KB, with the same frequency. This experiment measures throughputs and delays of these small transfers from node 2 to node 1.

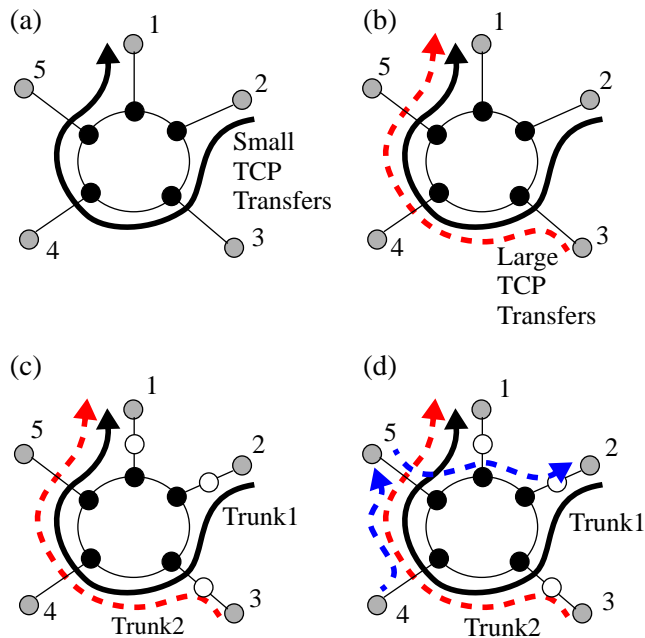

Figure 17. TCP Trunking Experiments Suite TT3 (a): small TCP transfers compete with large TCP transfers. Performance results are summarized in Table 1.

Case (a) of Figure 17 has only small TCP transfers with no competing traffic. In case (b), there are ten greedy long-lived TCP transfers from node 3 to node 1. In case (c), there are two trunks: one trunk carries small transfers from node 2 to node 1, and the other carries greedy long-lived TCP transfers from node 3 to node 1. In case (d), there are two additional greedy long-lived TCP transfers from node 4 to node 5, and from node 5 to node 2. Thus, in this case, there are multiple bottlenecks on the ring for small transfers from node 2 to node 1.

Table 1 shows average throughput and delay statistics for the small file transfers requested by the ten ftp clients at node 1. The experimental results show that these small transfers suffer when they compete with long-lived greedy TCP transfers. Their throughput is reduced from about 380 KByte/s to about 50 KByte/s. In the meantime, their mean, standard deviation, and maximum delay are increased. With TCP trunks, the situation is much improved. The throughput for web transfers increases from 50 KByte/s to about 215 KByte/s. The delay statistics are also substantially improved.

| Case | Average Throughput (KByte/s) | Delay Statistics (ms) for 8 K transfers | | |
|---|---|---|---|---|
| | | Mean | SD | Max |
| (a) | 380.05 | 451.5 | 147.9 | 1336 |
| (b) | 50.09 | 2183.0 | 1861.4 | 4182 |
| (c) | 215.73 | 719.1 | 262.2 | 1871 |
| (d) | 202.39 | 910.9 | 178.3 | 1997 |

Table 1. Performance results of TCP Trunking Experiments Suite TT3 (a) of Figure 17. Average throughputs and delays for small TCP transfers from node 2 to node 1 are much improved when TCP trunks are used.

**Experiment TT3 (b): Use of Trunks to Protect TCP Flows against UDP Flows**

This experiment, as depicted in Figure 18, has a set-up similar to that for Experiment TT3 (a). Case (a) has only small TCP transfers with no competing traffic. In case (b), there is a competing UDP flow from node 3 to node 4. This UDP flow is an on-off UDP flow with each on or off period lasting 10 ms. The source of the UDP flow tries to send as many 1024-byte UDP packets as possible during each on period. In case (c) there are two trunks: one trunk carries small file transfers from node 2 to node 1, and the other carries UDP traffic from node 3 to node 4. In case (d), there are two additional greedy long-lived TCP transfers from node 4 to node 5, and from node 5 to node 2.

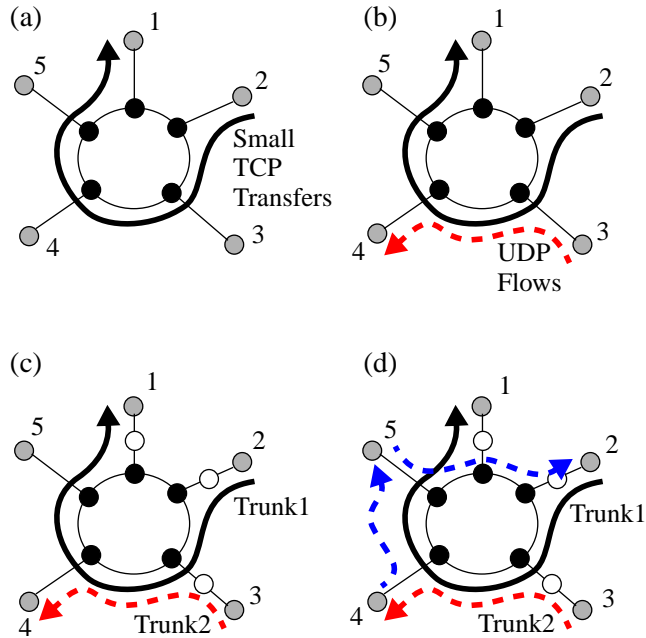

Figure 18. TCP Trunking Experiments Suite TT3 (b): small TCP transfers compete with UDP flows. Performance results are summarized in Table 2.

Table 2 shows average throughput and delay statistics for the small file transfers from node 2 to node 1. The experimental results show that these small transfers suffer when they compete with UDP traffic. Their throughput is reduced from about 380 KByte/s to about 53 KByte/s. Their mean, standard deviation, and maximum delay are increased. With TCP trunks, the situation is much improved. The throughput for small transfers increases to about 270 or 252 KByte/s for case (c) or (d), respectively. The delay statistics are also improved.

| Case | Average Throughput (KByte/s) | Delay Statistics (ms) for 8 K transfers | | |
|---|---|---|---|---|
| | | Mean | SD | Max |
| (a) | 380.05 | 451.5 | 147.9 | 1336 |
| (b) | 53.21 | 2541.1 | 4021.7 | 13053 |
| (c) | 270.45 | 507.9 | 136.5 | 1921 |
| (d) | 252.65 | 663.9 | 166.9 | 1892 |

Table 2. Performance results of TCP Trunking Experiments Suite TT3 (b) of Figure 18. Average throughputs and delays for small TCP transfers from node 2 to node 1 are much improved when TCP trunks are used.

**Experiment TT3 (c): Use of Trunks to Protect Sites with a Small Number of TCP Connections**

This experiment, as depicted in Figure 19, has two cases. In the first case, there are 5 greedy TCP connections from node 2 to node 1, and 15 greedy TCP connections from node 3 to node 1. In the second case, the two numbers of TCP connections are 15 and 45, rather than 5 and 15. In both cases, there are two additional greedy TCP connections from node 4 to node 5, and from node 5 to node 2.

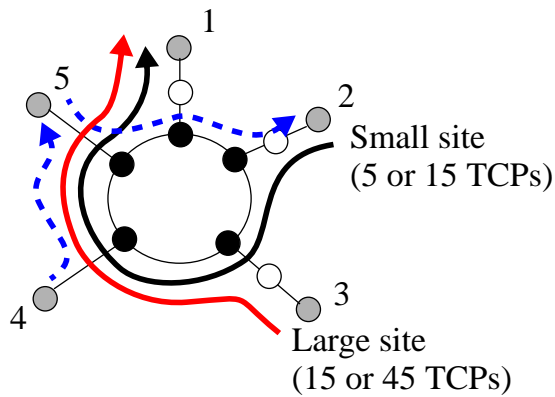

Figure 19. TCP Trunking Experiments Suite TT3 (c): a small site with a small number of TCP connections competes with a large site with a large number of TCP connections. Performance results are summarized in Table 3.

Table 3 shows the utilization of the link from node 2 or node 3 to the ring. Experimental results show that without trunks, the link utilization for node 2 that has a small number of TCP flows is much lower than that of node 3 that has a large number of TCP flows. When two TCP trunks are used from node 2 to node 1 and from node 3 to node 1, the unfairness problem is basically eliminated.

| # TCPs at node 2 / # TCPs at node 3 | Link Utilization (%) | | | |
|---|---|---|---|---|
| | without Trunking | | with Trunking | |
| | Node 2 | Node 3 | Node 2 | Node 3 |
| Case 1: 5/15 | 25.78 | 68.77 | 48.04 | 43.35 |
| Case 2: 15/45 | 4.62 | 85.09 | 47.29 | 46.14 |

Table 3. Performance results of TCP Trunking Experiments Suite TT3 (c) of Figure 19. TCP trunking substantially reduces the disparity between the utilization or throughput achieved by node 2 and node 3.

## 3.8 TCP Trunks Management and Migration Issues

### 3.8.1 Managing TCP Trunks

A TCP trunk may be set up for a given layer-2 circuit or an MPLS label-switched path. Setting up a trunk is simple and requires only setting up one control TCP connection and then changing a routing entry at the trunk sender to redirect certain user packets to the tunnel queue associated with the TCP trunk. Running TCP trunk senders or receivers on an edge router causes little processing overhead because all receiving and sending operations in the TCP decoupling approach are performed in the kernel. To support N TCP trunk senders or receivers, an edge router only needs to maintain N TCP control blocks in the kernel. The TCP processing module of the kernel is shared by these N TCP trunks.

### 3.8.2 Use of "Public Trunks" during Migration

Use of public trunks will help solve the problem that a site which first adopts TCP trunking will receive less bandwidth than its fair share when competing with other sites which have not adopted TCP trunking. This problem may occur because the site which aggregates many of its flows into a TCP trunk will have a decreased number of TCP flows that will compete with flows of other sites. As demonstrated in Experiment TT3 above, this site will likely to receive less bandwidth than its fair share.

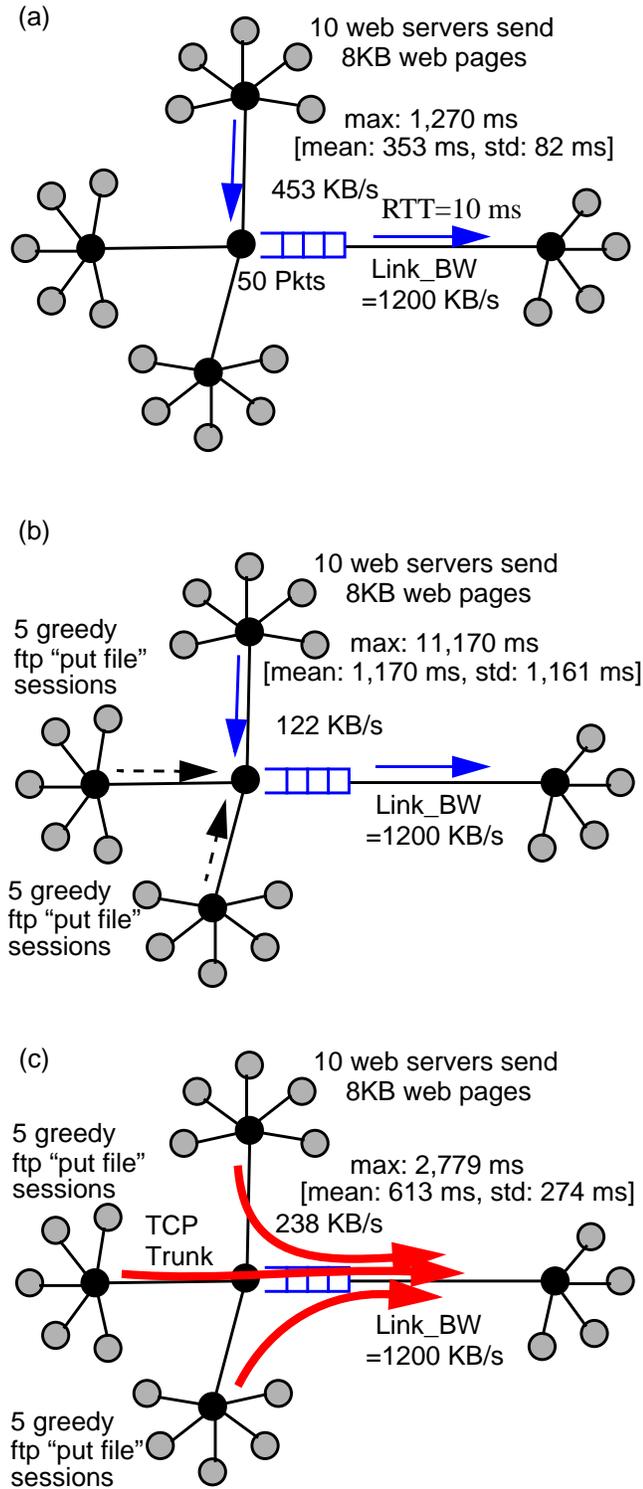

Figure 15. TCP Trunking Experiments Suite TT2. Web site throughput: (a) under no competing ftp traffic and (b) under competing ftp traffic. (c) Web side performance for load (b) when three TCP trunks, one for each site, are used.

To solve this problem and provide a migration path, the ISP of a domain can set up a few public TCP trunks and aggregate user packets of those sites, which have not yet adopted TCP trunking, into these public trunks. By properly configuring these public trunks' bandwidths, the ISP can make the total bandwidth available for flows of a site, which has not adopted TCP trunking, to be slightly less than the site's fair share. This policy will provide incentives for a site to adopt TCP trunking. Note that, no matter whether a site wants or not, technically the ISP can always enforce this policy by forcing incoming traffic to be aggregated into TCP trunks on all of its ingress routers.

### 3.8.3 Dealing with Routers that Do Not Support the Decoupling Packet Dropping Algorithm

During the migration process, some routers may not use the packet dropping algorithm discussed in Section 2.2.7 for TCP trunks implemented using the TCP decoupling approach. For example, some routers may still use a simple FIFO queue and scheduling algorithm or the RED packet dropping algorithm, and thus cannot distinguish between control and user packets. It is interesting to see how TCP trunks will perform in this situation.

Some experiments have been done to study this problem. The results show that, with the exception of the lossless property, all other properties of TCP trunks discussed in Section 3.5 still hold. In particular, using different VMSS values to allocate bandwidth in a fine-grain way as presented in Section 3.7 still works. This reason is that, despite the fact that a TCP trunk's user packets may be dropped now, the control TCPs of every competing TCP trunk still experience the same header packet drop rate and therefore achieves the same bandwidth for their header packets.

### 3.8.4 Non-Decoupling TCP Connections Compete with Decoupling TCP Connections

Sometimes non-decoupling (normal) TCP connections may compete with decoupling TCP connections (i.e., control TCPs and its associated data packet streams) for available bandwidth on a router which implements the decoupling packet dropping algorithm. This situation may happen as, in a domain, there may be a few TCP connections set up for management or control purposes (e.g., BGP [5] uses TCP to exchange routing information between edge routers in a domain). It is desirable that they can compete fairly with each other.

When the packet dropping algorithm discussed in Section 2.2.7 is used in router to first drop header packets when congestion occurs, it is obvious that non-decoupling TCP connections will receive an unfair advantage in sharing available bandwidth because none of its packets is explicitly marked as a "header" packet. To ensure the fairness, every packet of non-decoupling TCP connections should be marked as a "header" packet. This method is correct as a normal TCP packet contains both a header packet and its TCP payload. However, to be more practical while being able to achieve the same effect, instead, every user packet carried by a TCP trunk is explicitly marked as a "user" packet and the packets of non-decoupling TCP connections need not be marked. Any packet which is not explicitly marked as a "user" packet is considered as a header packet by default.

## 3.9 Comparison with Other Approaches

With a design goal to enforce fairness of achieved bandwidth between an application which uses a single TCP connection and an application which uses multiple parallel TCP connections to download web pages, some schemes have been proposed to aggregate many parallel TCP flows and put their aggregate transmission rate under a single TCP connection's congestion control.

Persistent connection HTTP (P-HTTP), proposed in [61], multiplexes multiple parallel TCP connection's traffic into one user-level persistent TCP connection between a web client and a web server to reduce web page download delays and put the aggregate transmission rate of these parallel TCP connection's traffic under a single TCP connection's congestion control. P-HTTP is essentially the same as the user-level implementation of TCP trunks and thus has the problems discussed in Section 3.4.1.

The approaches proposed in [29] modifies the TCP/IP stack so that a set of TCP connections share and use only one TCP connection's congestion control state variables. In particular, the set of TCP connections share a congestion window size variable. When any TCP connection experiences packet losses, the shared congestion window size is reduced by a half. When no TCP connection in a set experiences packet loss, the shared congestion window size is increased by one packet every RTT. Although the goal is to try to let the set of TCP connections receives the same treatment as a single TCP connection would receive, the modified TCP fast retransmit and fast recovery mechanism are no longer the same as the normal ones. Without careful verifications and tests, the designed TCP-like congestion control may be harmful to network congestion control. Also, the design and implementation are complicated.

The approach proposed in [30] implements a TCP-like congestion control algorithm between a sending and receiving nodes. The sending node sends probe packets to the receiving node and the receiving node sends back probe-reply packets. Depending on whether these probe packets are lost or not, the sending node uses TCP congestion control's principle (additive-increase and multiplicative-decrease) to maintain a TCP congestion window size. All traffic from the sending node to the receiving node are sent under the control of the estimated TCP congestion window size.

One possible approach to achieving the same effects of using trunks is to implement weighted round-robin packet scheduling on aggregate flows (a VP or a label-switched path) in every router in a network. Although this approach can achieve the same effects as using trunks, this approach requires weighted round-robin packet schedulers to be installed in every router in the core network and thus has deployment issues. Also as argued in [56], weighted round-robin packet scheduling alone can not eliminate the congestion collapse problem. Instead, end-to-end congestion control is still needed. Therefore, in the TCP decoupling approach, although weighted round-robin packet scheduling can be optionally used in the routers, control TCPs' using TCP congestion control for implementing end-to-end congestion control is still necessary.

Although ECN [53] has the potential to achieve the "lossless" property by marking packets in the routers when incipient congestion occurs, transporting a packet stream on top of an ECN-enabled TCP connection forces the packet stream to be subject to the problems caused by TCP's error control. In addition, ECN still has some open issues about its implementation. For example, when the TCP receiver echoes back the ECN bit, should it send back duplicate acknowledgement packets? If yes, when should it stop doing this? When a sender receives each ECN acknowledgement packet, should the sender reduce its congestion window size by half every time? What will happen if an ECN acknowledgement packet gets lost?

The proposed TCP trunk approach in this thesis is as natural as P-HTTP but does not have the problems with P-HTTP. Not like [29, 30], this approach implements a genuine TCP congestion control between the sending node and receiving node and thus can claim that the approach is 100% TCP, not just TCP-like. This 100% TCP property makes sure that the proposed TCP trunk approach will not do any harm to network congestion control. In contrast, an unverified TCP-like congestion control may increase TCP's aggressiveness or make the network congestion control unstable. Unlike the approach of using weighted round-robin scheduling in *every* router in a network, the TCP trunk approach is an edge-to-edge approach, which makes the deployment of TCP trunks much easier. Routers in the TCP trunk approach can use a simple and low implementation cost FIFO buffer and a RED-like packet dropping algorithm (discussed in Section 2.2.7) to support very high-speed links such as OC-192.

# Chapter 4  Application 2: Reliable Decoupling Socket for Wireless Communication

This section describes an application of the TCP decoupling approach to implement of a new kind of reliable socket services on hosts. This new kind of socket, called *reliable decoupling socket*, is suitable for reliable transport of data over networks which include wireless links that may corrupt packets due to link errors. Being able to manage TCP's congestion and error control separately, the reliable decoupling socket approach allows these new TCP-based services to make efficient use of error-prone wireless links' bandwidth. The approach can be especially useful for applications involving large RTTs, as in the case of satellite communications.

## 4.1 Introduction

Due to TCP's congestion control design (described in Section 2.1), when one packet of a TCP connection is lost, the sending rate of the TCP connection must be reduced by at least 50%. When multiple packets in a TCP's congestion window are lost, the TCP connection may time-out for more than one second. This congestion control design works well to prevent congestion in a network in which links (such as fiber optics) have very small bit-error-rates (BER) (e.g., $10^{-12}$) and therefore packet losses mostly result from packet dropping due to router buffer overflow during congestion. However, in a network in which links (such as wireless links) have large bit-error-rates (e.g., $10^{-6}$) and therefore packet losses may result from either packet dropping due to congestion or packet corruption due to link errors, TCP's congestion control design mistakenly and unnecessarily reduces a TCP connection's sending rate when its packets get corrupted and lost due to link errors. The result of the wrong control decisions is that the throughput of a TCP connection in such a network is very poor and a TCP connection cannot fully utilize the bandwidth of the wireless links.

Ideally, a TCP connection should be able to fully utilize a wireless network's bandwidth while at the same time avoid network congestion if TCP can distinguish packet losses caused by congestion from packet losses caused by corruption. Suppose that such a TCP connection exists. In order to fully utilize a wireless network's bandwidth and avoid congestion, the congestion control at the TCP connection's sending node should perform the following operations. For packet losses caused by congestion, the TCP congestion control should reduce the TCP connection's sending rate to remove the current congestion. For packet losses caused by corruption, the TCP congestion control should not be invoked to reduce the TCP connection's sending rate. Instead, the lost packets should be retransmitted and the following data should be transmitted using the current sending rate allowed by TCP's congestion control. Since TCP's congestion control uses an additive-increase (when there is no congestion) and multiplicative decrease (when there is congestion) algorithm to control a TCP connection's sending rate, when a packet is corrupted and lost, actually the sending rate of the TCP connection should keep increasing until one of its packet is really lost due to congestion, rather than being reduced.

Since TCP performs poorly on wireless links, Improving TCP's performance on wireless links has been a challenge and an active research topic for many years. There are many approaches proposed to address this problem [35, 26, 13, 16, 27, 3, 28, 47, 45]. Section 4.2 will briefly review these approaches and compare them to the reliable decoupling socket approach.

The rest of this chapter is organized as follows. Section 4.2 presents some related work showing other approaches to improving TCP's throughput in a wireless network. Section 4.3 analyzes how a non-zero bit-error-rate limits a TCP connection's maximum achievable throughput. Section 4.4 presents the reliable decoupling socket approach and its implementation. Section 4.5 discusses the strategies that are currently used in the reliable decoupling socket approach's error control to improve a TCP connection's throughput. Section 4.6 discusses why the reliable decoupling approach can achieve a significant improvement on TCP's throughput. Section 4.7 presents experimental results and shows that the reliable decoupling approach outperforms TCP Reno and TCP SACK. Section 4.8 compares the reliable decoupling socket approach to other approaches. Section 4.9 discusses some future work that can further improve the reliable decoupling socket approach's performance.

## 4.2 Related Work

This section briefly summarizes some mechanisms that have been proposed to improve TCP performance over wireless links.

> **Link-Layer Schemes (e.g., [13, 16])**: Forward error correction (FEC) schemes can be used to reduce the effective BER of a wireless link at the expense of reduced bandwidth and a requirement for high processing power to encode and decode packets. Automatic Request-Repeat (ARQ) can be used to retransmit lost packets at the link layer to hide packet loss from the sender of a TCP connection at the expense of increased delay and delay variations, and introduced packet reordering. These two schemes can be combined to improve the quality of a wireless link.
>
> **Snoop Protocol (e.g., [27]):** If only the last hop to a mobile host is a wireless link, a TCP-aware agent can be run on the base station to snoop passing TCP packets to do some local controls. For example, by caching recent transmitted TCP packets sent to a mobile host and watching the returning acknowledgment packets sent back to the sender of a TCP connection, the snoop agent can quickly resend a cached copy of a lost packet to the mobile host if it observes more than three duplicate acknowledgment packets are sent back to the sender of a TCP connection. This kind of scheme has some drawbacks as follows. First, the agent must be TCP-aware. Second, the snooping performance overhead is high. Third, although a lost packet can be retransmitted locally by the base station, the generated three duplicate acknowledgment packets still reach the sender of the TCP connection and cause the sender to unnecessarily reduce its sending rate by 50%.

**Split Connection (e.g., [3])**: If only the last hop to a mobile host is a wireless link, a TCP connection to a mobile host can be split into two connections. The first one starts at the sender of the TCP connection and ends at the base station. The second one starts at the base station and ends at the mobile host. Since the second TCP connection is explicitly used for the wireless link where packet losses are solely due to corruption, not congestion, it can be fined tuned to improve TCP performance on the wireless link. One drawback of this kind of scheme is that the end-to-end semantic of TCP is violated.

**Explicit Loss Notification (e.g., [28, 47])**: Like the snooping scheme, a TCP-aware agent is run on the base station to watch passing TCP packets to deduce that there may be a packet lost due to corruption. It then sets a special bit in the returning acknowledgment packets to notify the sender of a TCP connection that the recent packet loss may be a result of corruption, not congestion. When detecting this bit, the sender will not reduce its sending rate by 50%. The effectiveness of this kind of scheme depends on the correctness of the inferences. This kind of scheme is an improved version of snoop protocols but still has some drawbacks as follows. First, the agent must be TCP-aware. Second, the snooping performance overhead is high. Third, the TCP congestion control at the sender of a TCP connection needs to be modified and becomes more complicated.

**Standard TCP Mechanisms (e.g., [45])**: TCP SACK can be used to recover from multiple packet losses in a window without timing-out. The same TCP congestion control algorithms, but different parameters, can be used to more aggressively transmit data. For example, "ack every other packet" can be changed to "ack every packet" to increase a TCP connection's ramp up speed. One drawback of this kind of scheme is that these modifications to TCP may result in a too aggressive TCP protocol, which is harmful to congestion control in a network.

## 4.3 An Analysis of the Effect of Bit-Error-Rate on a TCP Connection's Maximum Throughput

This section presents an analysis showing the effect of a non-zero BER on the maximum bandwidth an idealized TCP connection can achieve if the idealized TCP cannot distinguish packet losses caused by congestion from those caused by corruption. An idealized TCP connection is defined as a TCP connection whose fast retransmit and recovery mechanism always works on packet losses and never times-out.

Suppose that the Bit Error Rate (BER) of a wireless link involved is $3*10^{-5}$, which is a typical value [4]. Assume that, just for this analysis, the packet size PS is a typical MTU of 576 bytes, and that bit errors are randomly distributed in packets. (Many researchers used this model to model bit errors caused by additive white Gaussian noises. See [4] for an example.) Then the Packet Error Rate is close to:

$$PER = BER*PS = 3*10\text{-}5*576*8 = 0.14 \qquad (3)$$

This means that on average one corrupted packet is expected to occur in every 1/PER = 7.2 packets.

The following calculates, for an idealized TCP connection under such a PER, its the Maximum Allowable Throughput (MAT) in bits per second, and the Maximum Allowable Window (W) in packets. These numbers are achieved when there is no TCP time out, otherwise MAT and W would be smaller. That is, this session assumes that a corrupted packet can always be recovered by fast retransmit and recovery mechanism. Since when a packet is lost, TCP's congestion control will cut its current congestion window size W by a half to become W/2, and then increase the congestion window size by one packet every the TCP connection's round-trip time (RTT) until one packet is lost again, the number of packets transmitted between two packet losses is thus W/2 + (W/2 + 1) + ... + N, where N is the window size when the next corrupted packet occurs. Furthermore, since after the next corrupted packet occurs, the current congestion window size N needs to be cut to N/2 as well and the same cycle (N/2 + (N/2 + 1) ...) repeats, in order for the cycle to repeat forever so that the system is dynamically stable, W/2 must be the same as N/2, which means that N is equal to W. As a result, in each cycle, the TCP connection's window size will grow from W/2, (W/2)+1,...., to (W/2 + W/2), and the total number of packets transmitted between two packet losses is W/2 + (W/2 + 1) + ... + (W/2 + W/2) = (3/8)W*W + 3W/4.

Therefore,

1/PER
= W/2 + (W/2 + 1) + ... + (W/2 + W/2)
= (3/8)W*W + 3W/4 (4)

Based on Equation (4), given a PER value, W can be solved. For example, when PER is 0.14, W is about 4.

Note that the congestion window grows by one packet per RTT. Thus a total of (3/8)W*W + 3W/4 packets are sent over (W/2)*RTT time as depicted in Figure 20. This phenomenon implies that

MAT
= ((3/8)W*W + 3W/4)*PS*8 / ((W/2)*RTT)
= (3/4)*W*PS*8/RTT + (3/2)*PS*8/RTT (5)

where RTT is the TCP connection's end-to-end round-trip-time, in seconds, for the TCP connection.

Based on Equation (5), for example, when W = 4, RTT = 0.540 and PS = 576,

MAT = 26 kbps (6)

Using a retransmission packet loss detection algorithm, Samaraweera and Fairhurst [47] report that their method can achieve an optimal throughput of about 26 kbps under similar assumptions about BER, packet size and RTT. The matching of the above analytic results with their empirical results provides a validation for the above analytical reasoning.

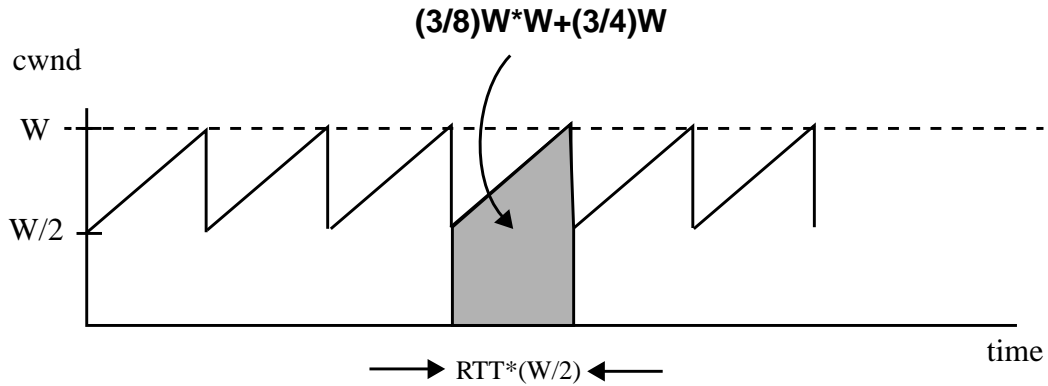

Figure 20. TCP's saw-tooth window growing and shrinking behavior

The current MAT bound of Equation (5) or (6) results from link errors rather than network congestion. The bound will hold even when there is no congestion in the network and the link bandwidths are infinitely large. The severity of the problem increases when RTT is large, as in the case of satellite communications [35, 45]. This poor TCP throughput presented in the analytical result is a consequence of wrongly applying TCP congestion control algorithms to a situation where packet losses are due to link errors, rather than due to congestion.

## 4.4 The Reliable Decoupling Socket Approach and Implementation

The reliable decoupling socket approach is a direct application of the basic TCP decoupling approach. The reliable decoupling socket approach applies TCP's congestion independently from TCP's error control to a stream of data packets. In this approach, a TCP connection is set up as normal between the sending and the receiving hosts to reliably transport data from the sending host to the receiving host. It is called "data TCP connection", or more briefly, "data TCP" in this thesis as its function is to solely transport data. A TCP circuit is then set up between the same sending and the receiving hosts. The data packet stream generated by the sender of the data TCP is then sent into the TCP circuit. The sender and receiver of the data TCP handle only error control. Their TCP congestion control is disabled and the data packets generated by the sender of the data TCP can be sent into the TCP circuit at the maximum speed allowed by the TCP circuit (i.e., as long as the tunnel queue of the TCP circuit is not full). The TCP circuit uses TCP congestion control to probe for available bandwidth in networks via its tiny header packets. Its congestion control is triggered only when its tiny header packets are corrupted and lost; a corrupted and lost data packet whose transmission rate is regulated by the TCP circuit will not trigger the TCP circuit's congestion control. Because the PER of these tiny header packets is much smaller than that of full-size packets carrying MTU data payload, the probability of mistakenly triggering TCP congestion control to reduce the sending rate upon packet corruption is significant reduced. The reliable decoupling socket approach thus can provide a reliable and high-throughput data transfer in wireless network while using TCP congestion control to avoid network congestion.

The established TCP circuit can be just one control TCP. Figure 21 depicts the internal implementation of the two reliable decoupling sockets residing on the sending and receiving hosts. Since the TCP circuit uses only one control TCP, the reliable decoupling socket on each of the sending and receiving hosts is implemented internally as two TCP sockets -- one control and one data sockets. The control socket is associated with the control TCP. The data socket is associated with the data TCP, on which application program's data are transmitted. Following the decoupling principle, data packets will be sent at rates under the control TCP's congestion control. The data socket is provided to the user for transmitting the user's data whereas the control socket is hidden and invisible to the user.

While the control TCP sets sending rates for data packets, the data TCP is responsible for retransmitting corrupted or lost application data. The data TCP makes direct use of TCP's existing facilities such as sequence numbers and triggering mechanisms for packet retransmission. The data TCP does not deal with congestion control; its congestion window size (cwnd) is always set to infinite, except when a lost packet needs to be retransmitted. When retransmitting a lost packet, the data TCP will temporarily set the congestion window size (cwnd) to one MSS so only one packet is retransmitted. After retransmitting the lost packet, the cwnd is reset to infinite. At the sender of the data TCP, outgoing data packets are redirected and sent to the tunnel queue of the TCP circuit. The sender can send its data packet to the tunnel queue as fast as it can as long as the tunnel queue does not overflow.

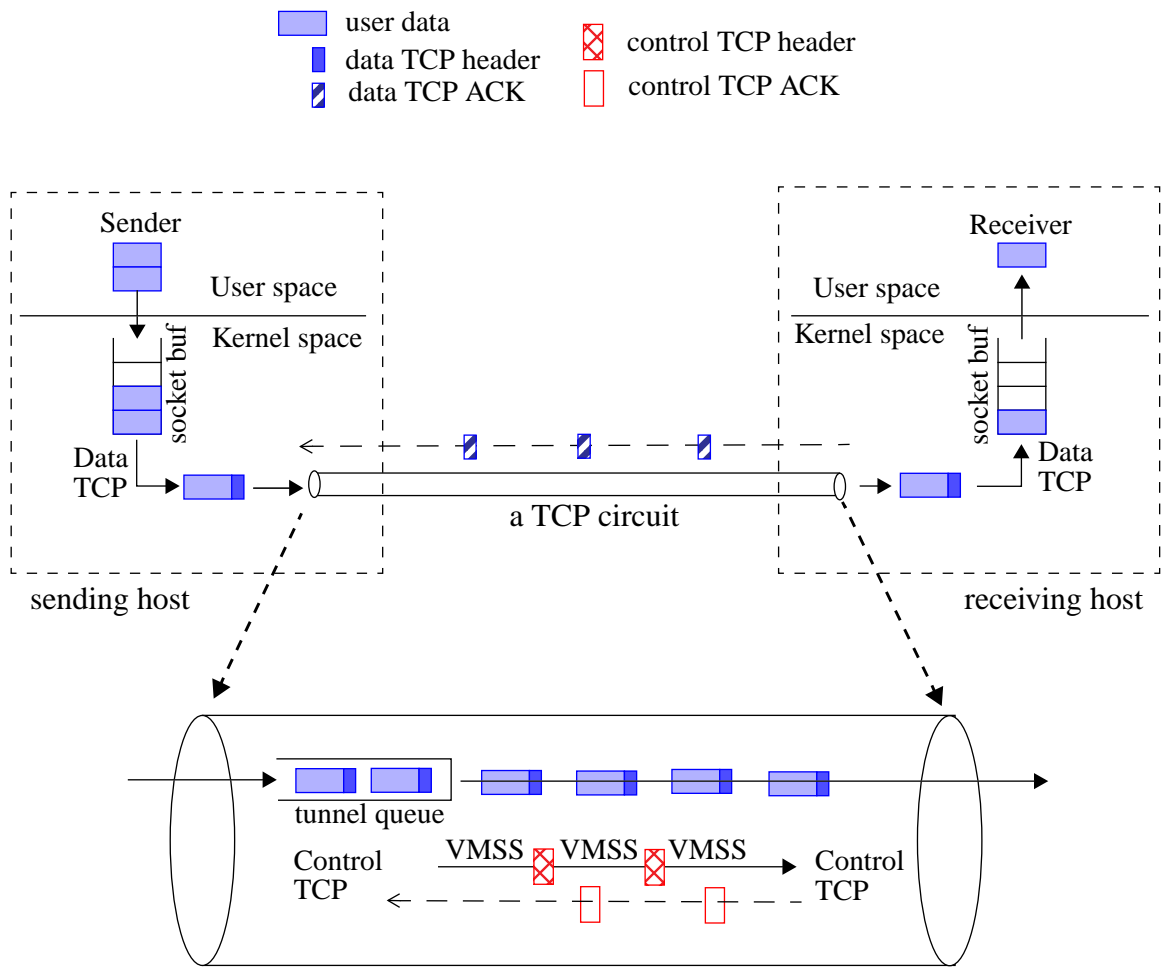

Figure 21. The internal implementation of the reliable decoupling socket on hosts.

## 4.5 Discussions on Error Control

Experimental results on testbed networks show that it is important for the data TCP to be aggressive in retransmitting lost data, as long as their sending is allowed by the congestion window of the control TCP. Otherwise, timeouts on the data TCP could happen easily, and performance can degrade drastically. To achieve the high goodputs reported in Section 4.7, the data TCP in the current implementation has the following features:

F1. The receiver uses the SACK option [43] to report to the sender up to three missing packets in an acknowledgment packet.

F2. The sender retransmits the first unacknowledged packet every time when some number X of additional duplicate acknowledgment packets are received. The number X is the current window size of the control TCP. Thus the method will retransmit again a previously retransmitted packet should it get corrupted or lost. This method can minimize chances of timeout.

F3. The sender uses a fine-grain retransmission timer of 50ms, rather than the system default of 500 ms, and disables the timer's exponential backoff.

Features F1 and F2 greatly reduce possible timeouts of the data TCP. Should timeouts still happen, F3 will minimize the negative impacts of time-outs on performance.

It is important to emphasize that the data TCP will send applications data under these aggressive send features, only when the sending is allowed by the congestion window of the control TCP. In the reliable decoupling socket implementation, the control TCP uses TCP Reno, a normal congestion control algorithm, with a coarse-grain retransmit timer of 500 ms, exponential backoff enabled, and the normal 3-duplicate acknowledgment packets trigger of packet retransmission. That is, the control TCP does not employ any aggressive feature such as F1, F1 and F3 at all. Since the control TCP is not aggressive and it controls the sending rate of the data TCP, the use of these aggressive retransmission features by the data TCP causes no harm to other network users.

However, the data TCP should not be unnecessarily aggressive. Otherwise, retransmission may become excessive and will hurt the overall goodput of the reliable decoupling socket. For feature F1 above, the number X is linked to the current window size of the control TCP to reduce the chance of premature retransmission due to an unnecessarily small X.

## 4.6 Why the Reliable Decoupling Socket Approach Improves TCP Performance

Since the reliable decoupling socket approach is a direct application of the TCP decoupling approach and thus has all of the TCP decoupling approach's properties, the following will sometimes use the name "TCP decoupling approach" for the name "the reliable decoupling socket approach" when analyzing and reporting performance. The reason is that the performance gain achieved by the reliable decoupling socket approach actually relies on the TCP decoupling's fundamental properties.

In the reliable decoupling socket approach, as shown in Figure 21, it is the control TCP that controls the sending rate of the data TCP's packets. The data TCP uses only TCP error control, but not TCP congestion control, to retransmit lost data packets or to transmit data packets as fast as the control TCP allows. Since the header packets, which are sent by the control TCP, are now the only packets whose losses will trigger TCP's congestion control to reduce the data TCP's sending rate, their small packet length of only 52 bytes in the TCP decoupling approach significantly reduces the chances of ambiguity of packet dropping and packet corruption for TCP's congestion control mechanism. In [13, 14], experimental results supported that, as packet size becomes smaller, packet error rate also becomes smaller.

An analysis for the TCP decoupling approach, which is similar to the analysis presented in Section 4.3 for a normal TCP connection, is presented as follows. Since it is the corrupted header packets rather than the data packets that will cause TCP congestion window size reductions, the size of header packets, rather than the combined size of both a header and data payload, should be used in computing W. When using a packet size of 52, instead of 576 used earlier, the computed W now becomes 14 instead of 4. Computing MAT using this new value of W = 14 and the original packet size of 576 results in a new

MAT of 119 kbps rather than its old value of 26 kbps in Equation (6) -- a speed up of 119/26 = 4.57!

The increase of W from 4 to 14 is more significant than just an increased MAT. A window size around 4 packets is hardly sufficient for supporting the fast retransmit and recovery mechanism since the mechanism relies on receiving three duplicate acknowledgment packets to trigger the retransmission of a lost packet. If fast retransmit and recovery mechanism is usually not triggered, the TCP connection will experience frequent timeouts. These TCP timeouts will severely impair TCP's performance in throughput, delay and fairness. An increase of the window size to a sufficiently large value such as 14 eliminates this timeout problem.

According to Equation (4), the TCP decoupling approach achieves a performance speedup proportional to sqrt(**MTU**/**HP_Sz**) over the normal TCP approach. It is obvious that if **HP_Sz** can be further reduced**,** the TCP decoupling approach will achieve an even higher performance speedup. Actually, it is the effective PER of header packets that matters as the ultimate goal is to reduce the effective PER of header packets to zero so that no congestion control will be wrongly triggered. Section 4.9 will discuss some schemes that can either physically reduce **HP_Sz** or reduce the effective PER of header packets.

## 4.7 Experimental Results

**Descriptions of Experiments**

On the testbed network described in Figure 11, there are two user TCP connections (one is from node A to G, the other is from node C to H) contending for the bandwidth of a wireless link, which is the link between node E and F and was simulated by an Ethernet. The experiments use an Ethernet link to simulate a wireless link, rather than directly using a real wireless link such as a WaveLAN network [65], because the experiments need to precisely generate and control the desired BERs, and precisely generating and controlling the desired BERs are hard to achieve using a real wireless link. Besides, WaveLAN implements IEEE 802.11 protocol [4] and thus employs ARQ to retransmit a corrupted packet up to 4 times. Because the experiments wants to clearly identify and evaluate the TCP decoupling approach's performance without ARQ's interference, the experiments did not use WaveLAN.

As described in Section 4.4, each of these two contending connections is internally implemented as a pair of data and control TCPs. Both the control and data TCPs use TCP Reno and the data TCPs are enhanced with features F1, F2 and F3 of Section 4.5. Performance numbers on TCP SACK are obtained from hosts running Window 98, which has a built-in version of TCP SACK. The experiments focus on the aggregate goodputs (measured at the application layer) of these two connections under varying BERs and RTTs on the simulated wireless link. In order to generate a given BER, bit errors were randomly generated on the simulated wireless link according to the given BER [4]. The size of the buffer from which data are transmitted to the simulated wireless link is 50 packets for header packets (it is HP_Th presented in Section 2.2.7). The total buffer size for both header and data packets is provisioned based on the ***Required_BS*** of Equation (2) of Section 2.2.7.2. The size of data packets is 1500 bytes (Ethernet's MTU) and the size of the header packets is 52 bytes (40 bytes TCP/IP header + 12 bytes TCP timestamp option).

**Reliable Decoupling Socket Experiments Suite RS1:**

This experiments suite demonstrates that the reliable decoupling socket implementation, called TCP Decoupling, can generally achieve much higher goodputs than TCP Reno and TCP SACK, for BER ranging from $10^{-7}$ to $10^{-5}$. With RTT = 10 ms, the results are summarized in Figure 22.

The top curve in Figure 22 is a theoretical upper bound on the goodput that the TCP decoupling scheme (or any other scheme) can possibly achieve, over a 10 Mbps lossy link. This curve is obtained by using goodput = maximum_link_goodput * (1 - packet_ error_rate). Since packet_error_rate, i.e., PER, must increase as BER increases, the theoretical goodput upper bound must decrease as BER increases. In this idealized design and implementation, packet retransmissions take no time and there is no redundancy in retransmitting lost packets.

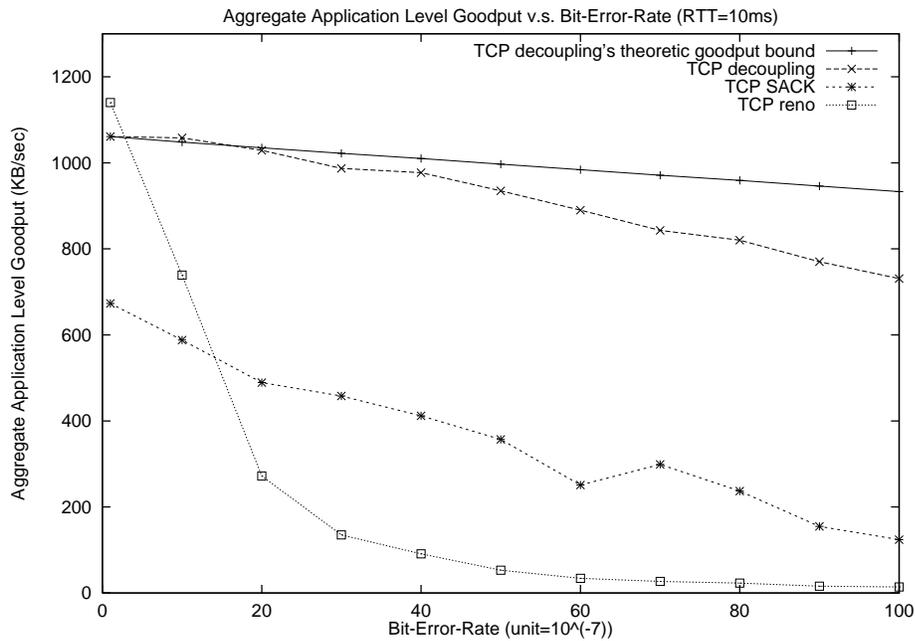

Figure 22. Performance improvements of TCP decoupling compared to TCP Reno and TCP SACK for various values of BER. RTT = 10 ms.

There is a gap between the curve of the theoretical upper bound and that of TCP decoupling. This is due to some unnecessary retransmissions in the TCP decoupling scheme, as discussed in the end of Section 4.5. When BER is near $10^{-7}$, TCP reno's goodput is slightly higher than that of TCP decoupling. The difference is due to the approximately 3% header packet overhead in this particular implementation of TCP decoupling. Note that if throughputs instead of goodputs were plotted, TCP decoupling's throughput still remains high even though BER increases.

**Reliable Decoupling Socket Experiments Suite RS2:**

Experiments Suite RS2 is similar to RS1 above, but with varying RTTs. A fixed value of BER = $2*10^{-6}$ is used in these experiments. Figure 23 shows that TCP decoupling always outperforms TCP reno and TCP SACK.

The top curve in Figure 23 gives a theoretical upper bound on the goodput that the TCP decoupling scheme can possibly achieve for various values of RTT. The declining trend of the upper bound as RTT increases, depicted in Figure 23, is an inevitable consequence of BER > 0. Equations (3) and (4) show that W is a function of BER and PS. (In fact, W is inversely proportional to the square root of BER*PS.) For the experimental suite RS2, since BER and PS are fixed, so is W. Equation (4) shows that for a fixed W, MAT must decrease linearly as RTT increases. Figure 23 shows that the achieved goodput of the TCP decoupling scheme approaches its upper bound, although it does not match due to some retransmission redundancy.

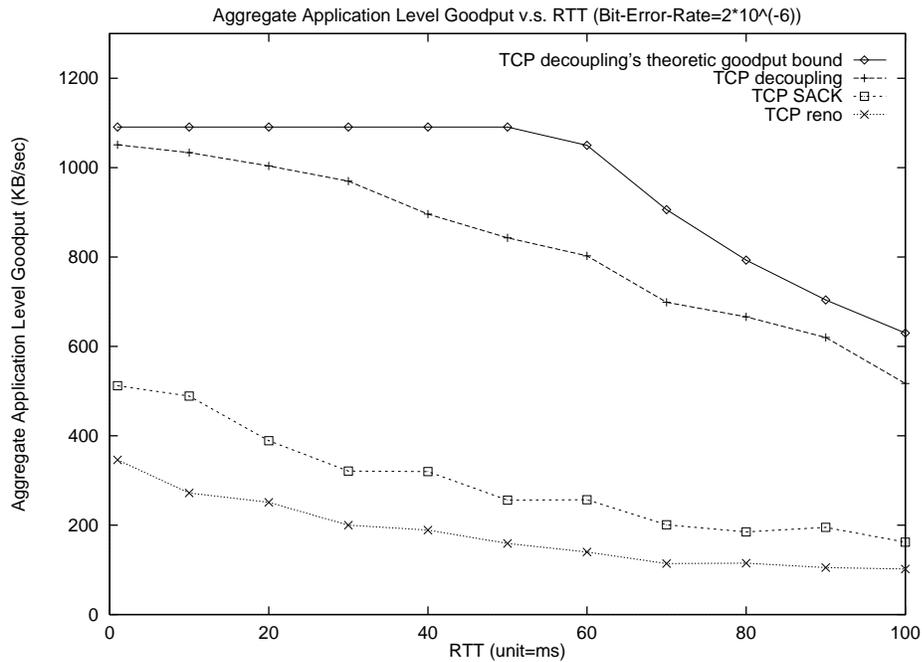

Figure 23.    TCP Reno and TCP S

Figure 23 shows that the goodput of TCP decoupling at RTT = 100 ms is approximately 560 KB/sec. This goodput is close to the best possible performance under the TCP decoupling approach. With BER = $2*10^{-6}$ and PS = 52 bytes for header packets, Equation (3) implies PER = 0.0008. For this value of PER, solving Equation (4) for W gives W = 56. With W = 56, PS = 1500 bytes for data packets, and RTT = 100 ms, Equation (5) gives MAT = 652,500 Bytes/sec. After accounting for the packet error rate of BER*1500*8 = 0.024 for data packets, and the overhead of the 52-byte TCP/IP header associated with each 1500-byte data packet, a theoretical upper bound on the goodput of approximately 615 KB/sec is obtained. This upper bound is about 9% higher than the achieved goodput of 560 KB/sec.

Several experiments using WaveLAN network cards were also conducted. The attempt to generate and control certain desired BERs by gradually increasing the distance between the sending and receiving nodes turned out to be unsuccessful. The measured results were very susceptible to the environment and were hard to reproduce. However, in some environments, the observed performance speedup of a TCP connection in the TCP decoupling approach over a normal TCP connection was about 3.7. This speedup is close to that predicted by Equation (5) when the MTU is 576 bytes, **HP_Sz** is 52 bytes, accounting for the overhead of header packets.

## 4.8 Comparison with Other Approaches

The reliable decoupling socket approach is an end-to-end approach. Unlike many other schemes presented in Section 4.2, it does not require a special TCP-aware agent to run on the base station to snoop passing TCP packets, nor does it need to split a TCP connection into two connections at the base station. The reliable decoupling socket approach does not need any support from a wireless network. Thus, the wireless network can be simple and easy to implement. Due to the reliable decoupling socket's end-to-end property, it can be quickly deployed in any kind of wireless network to receive its offered performance improvement. On the contrary, many of the other schemes presented in Section 4.2 have special demands for the wireless network, and thus cannot be used in any wireless network. For example, snoop and split schemes are not suitable to a multi-hop all-wireless network because it is impractical to snoop the traffic of a TCP connection or split a TCP connection multiple times on every router along the TCP connection's path.

## 4.9 Future Improvements

Equation (4) shows that it is advantageous to use tiny header packets to implement TCP congestion control so that the ambiguity of packet dropping and packet corruption can be greatly reduced. Currently, the size of a header packet of 52 bytes has reached the minimum for a packet to be an TCP/IP packet carrying the useful TCP timestamp option, which allows for a more accurate estimate of a TCP connection's RTT.

Using the TCP header compression algorithm proposed in [60] and the *twice* algorithm proposed in [46] on wireless links can greatly reduce the size of header packets (and thus their PERs) without the bad effects on TCP's performance caused by dropping a header-compressed packet [46]. The TCP header compression mechanism can compress the TCP/IP header of a header packet from 40 bytes down to only 3 bytes, resulting in a 3+12 (TCP timestamp option) = 15 byte packet. (Note that the TCP header compression algorithm does not attempt to compress TCP options. However, the same method can be used to also compress the TCP timestamp option and result in a packet size lower than 15 bytes.) *Twice* algorithm works with the TCP header decompresser at the receiving end of a wireless link. If the decompresser detects state inconsistency (by noticing the wrong computed TCP checksum) when decompressing a header-compressed packet, *twice* first assumes that a packet has been dropped and makes a guess of the content of the dropped packet's TCP/IP header based on the past history of TCP header contents. It then advances its decompression state as if the lost packet had been correctly received and decompressed, and then decompresses the newly arrived header-compressed packet again. If the computed TCP checksum is correct, the guess that one packet is dropped is correct and every thing is back to the normal state. Otherwise, *twice* assumes that two packets are lost and the above procedure repeats. It is worth noting that TCP header compression and *twice* are particularly well suited to the TCP decoupling approach as the difference between consecutive header packets are only in the sending sequence number field and the difference is always VMSS. This enables the TCP header compression to always compress a 40-byte header into a 3-byte header and makes *twice* very easy to make a correct guess. For *twice*, because the PER of the tiny header packets is further significantly reduced by TCP header compression, the probability that more than one consecutive header packets are dropped becomes exponentially-reduced small, which makes *twice* work in its first guess almost every time.

Another dimension of improvement is to apply FEC and/or ARQ to only header packets to protect them from corruption so that the effective PER of header packets is reduced. Since the size of header packets is small, the added bandwidth overhead caused by applying FEC to only these header packets is also tiny compared to the added overhead when FEC is universally applied to both the header and data payload of a 1500-byte TCP/IP packet in traditional approach.

Yet another dimension of improvement is to use a larger VMSS at the expense of generating more bursty traffic in networks. It is clear that due to a non-zero PER for header packets, there must be a limit on the bandwidth achieved by header packets. Since the achieved bandwidth of data packets is VMSS times that of header packets (discussed in Section 2.3.1), if the achieved bandwidth of data packets does not reach the wireless link's bandwidth, the VMSS can be increased to achieve a higher link utilization.

All of the schemes described above are currently under study. Their relative further improvement on TCP performance in the TCP decoupling approach will be published in the future.

# Chapter 5  Application 3: Unreliable Decoupling Socket for Multimedia Streaming

This chapter describes an application of the TCP decoupling approach to implement a new kind of unreliable socket services on hosts. This new kind of socket, called *unreliable decoupling socket*, is suitable for multimedia audio/video streaming applications.

## 5.1 Introduction

Today most streaming multimedia applications use UDP rather than TCP to transport a audio/video packet stream. The reasons why these applications don't use TCP include:

C1.  TCP may time out for more than one second without sending any data. A lengthy time-out is unacceptable for these applications as normally they need some minimum sending rate.

C2.  TCP's automatic retransmission of possibly lost packets is generally not desired, as retransmitted packets may arrive too late to be useful for these applications.

C3.  TCP may unnecessarily delay the delivery of arrived packets to the application on the receiving host. Due to TCP's insistence on providing a reliable and in-sequence delivery service, when there is any packet loss, packets that have already arrived may need to wait in the TCP assembly queue at the TCP receiver until the repair packet finally arrives.

However, since UDP itself does not automatically retransmit lost packets, UDP-based applications need to deal with the loss problem. Otherwise, the received audio/video frames will be broken and the perceived quality will be bad.

Another problem with using UDP is that UDP itself does not perform any congestion control. As a result, when sharing network bandwidth, UDP-based applications can have unfair advantages over TCP-based applications. Recently, how to make a UDP packet stream behave TCP-friendly when competing with TCP connections has become an important and active research topic [36, 15, 37, 41, 31, 52].

The TCP decoupling approach proposed in this thesis offers a new direction in implementing TCP-friendly protocols for multimedia streaming applications. It can make a UDP audio/video packet stream 100% TCP-friendly while removing all the TCP drawbacks outlined as C1, C2, and C3. When routers in a network can support the packet dropping scheme presented in Section 2.2.7 to first drop header packet before dropping data packets when congestion occurs, the TCP decoupling approach can guarantee that there is no data packet dropping due to congestion and every received frame is good (not broken).

## 5.2 Related Work

There are three common approaches to error control for multimedia streaming applications. The first approach is to retransmit lost packets if it is still not too late for the receiver to play them. The second approach is the use of Forward Error Correction (FEC) [13, 6]. The sender sends redundant information for each block of packets so that if only a few packets in the block are lost, the receiver can reconstruct the lost packets. The third approach uses Error Concealment methods [6, 22]. The receiver will try to reconstruct missing packets and mitigate their negative effects on the perceived quality to the user. Typical reconstruction methods include insertion of some templates and use of interpolation.

To give a UDP packet stream TCP's congestion control, in [36, 15, 37, 41, 31, 52], many congestion control schemes for UDP have been proposed. The design goal is to let a UDP flow achieve the same bandwidth as a TCP flow would achieve if TCP had been used to replace UDP for transporting user data in the same network environment. All of these schemes try to simulate TCP's congestion control algorithms. That is, they all maintain a congestion window size cwnd and update its value based on TCP's additive-increase and multiplicative-decrease principle upon the events of packet arrivals and packet dropping.

## 5.3 The Unreliable Decoupling Socket Approach and Implementation

The unreliable decoupling socket approach is a simple extension of the basic TCP decoupling approach. To use TCP congestion control to regulate the transmission rate of a UDP audio/video packet stream, the unreliable decoupling socket approach sets up a TCP circuit between the sending and receiving hosts of the UDP packet stream, and then pipes the UDP packet stream through it. This approach is very similar to the approach used for improving TCP throughput in wireless networks presented in Section 4.4. The only difference is that in wireless communication application, a TCP packet stream is regulated by TCP congestion control, however, in multimedia streaming application, a UDP packet stream instead is regulated by TCP congestion control. Since the control TCP of the TCP circuit uses normal TCP congestion control to regulate the sending rate of a UDP packet stream, the approach results in a 100% TCP friendly UDP packet stream.

The unreliable decoupling socket approach provides an unreliable transport-layer service on hosts, called unreliable decoupling socket, that will use UDP to transport application data while using TCP congestion control to adapt the UDP packet stream's bandwidth usage to network congestion. The implementation of the unreliable decoupling socket is parallel to, but simpler than, that of the reliable decoupling socket of Section 4.4. Unlike the implementation of Section 4.4, since multimedia data such as audio/video should not be automatically retransmitted when they are lost, a data UDP connection rather than a data TCP connection is used to transport data.

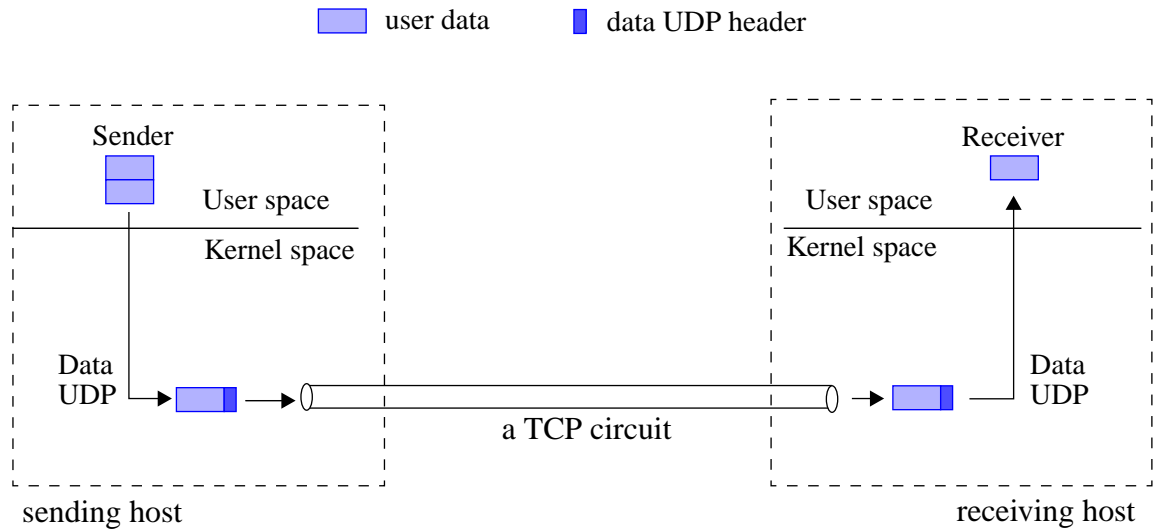

Figure 24. The internal implementation of a unreliable decoupling socket on the sending and receiving hosts.

More precisely, the unreliable decoupling socket is implemented internally as one UDP socket and one TCP circuit working together, as shown in Figure 24. Outgoing UDP packets are redirected into the TCP circuit so that their sending rate is regulated by TCP congestion control. The UDP socket is provided as normal to an application for transmitting applications data, but the TCP circuit is hidden and invisible to the user.

When UDP packets are inserted into a TCP circuit, to prevent them from being dropped due to buffer overflow at the tunnel queue of the TCP circuit, a new system call is provided which can be called by an application to report the current buffer occupancy of the tunnel queue. By providing this control feedback, the application in the sender will know whether it can send more data out without dropping them at the tunnel queue. If the routers on the UDP packet stream's routing path all support the packet dropping scheme designed for the TCP decoupling approach as described in Section 2.2.7, once an application can enqueue its data into the UDP socket without being dropped in the tunnel queue of a TCP circuit, the application can be assured that these packets will arrive at its receiver without being dropped in the network due to congestion. When a minimum rate is desired, the TCP circuit can be allocated a GMB so that the UDP packets in the tunnel queue can be forwarded out at this rate at the minimum. Creating an unreliable decoupling socket can be easily done in a user process by setting up the TCP circuit, and associating it with the UDP socket.

Since multimedia streaming packets are time-sensitive, a system call is provided for the sender application to manipulate packets that are already queued in the TCP circuit's tunnel queue. Therefore, if due to TCP congestion control, a packet in the tunnel queue has become too old to be useful, the sender can pull it out and replace it with the most recent audio/video frame packets.

## 5.4 Comparison with Other Approaches

Unreliable decoupling sockets described here are well suited for many multimedia streaming applications. The reasons are as follows.

First, by using unmodified genuine TCP congestion control for control TCP of a TCP circuit, it is guaranteed that the regulated streaming flow is 100% TCP-friendly. Although there are many proposed TCP-friendly schemes [36, 15, 37, 41, 31, 52, 38] developed based on theoretic models, using just a few simulations and experiments to validate their design goal, none of them can guarantee that its scheme will result in a 100% TCP-friendly UDP packet stream under any condition in a network. Maybe that is why more recently this kind of scheme is called "TCP-like" [41, 31].

Second, if the routers on the UDP packet stream's routing path all support the packet dropping scheme designed for the TCP decoupling approach, since the underlying decoupling network can guarantee that no application UDP packets will be dropped in the network due to congestion, a streaming UDP packet, once enqueued into the tunnel queue of a TCP circuit, will successfully arrive at its destination. As a result, every received frame at the receiver is a good frame as long as there are no link or node failures. There is no need to use FEC at the sender or error concealment methods at the receiver to deal with packet-loss problems caused by congestion. In this approach, the multimedia streaming sender can use the tunnel buffer occupancy of the TCP circuit as feedback to dynamically adjust its encoding rate, frame grab rate, resolution, and quantization level (similar to those suggested in [39]) to match the available bandwidth in the network. Therefore, all used network bandwidth is for good frames, and network resources are not wasted for packets that are dropped due to congestion.

Although retransmission, FEC, and error concealment methods can deal with the packet loss problem, all of these approaches come with expenses. For the retransmission scheme, since multimedia streaming applications normally have a tight delay bound on a packet's arrival time, if the end-to-end delay is large, there is no chance to wait one RTT and then use retransmission to recover lost packets. For the FEC scheme, network bandwidth is wasted for the inserted redundant information and unnecessary delay is introduced for collecting a block of packets before FEC encoding can start. Besides, packet reordering may be introduced if packet-level FEC is used. As for the error concealment method, the quality of reconstruction may not be good enough, and the implementation cost can be high [6, 22].

# Chapter 6  Conclusions

The TCP decoupling approach proposed in this thesis is a new, general, and powerful approach for applying TCP's congestion control alone to many application areas where TCP's other properties are not desired. This thesis has presented three such applications which are enabled or improved by the TCP decoupling approach.

The basic idea of decoupling control from data is not new. For example, the ATM ABR [1] scheme uses Resource Management Cells to establish proper sending rates for data cells. However, this thesis is the first to propose using a TCP connection's header packets and its congestion control to probe for available bandwidth for regulating the transmission rate of a packet stream. The TCP decoupling approach is able to take advantage of TCP's sophisticated congestion control without suffering from the problems caused by TCP's error control. This new approach opens up many opportunities. Several new and useful applications have been enabled and improved by the TCP decoupling approach. The TCP trunking, wireless communication, and multimedia streaming applications presented in this thesis are three of such successful examples. It is expected that many more additional applications that benefit from using the TCP decoupling approach will emerge in the future.

# Bibliography

[1] ATM Forum, "Traffic Management Specification 4.0"

[2] A. K. Parekh and R.G. Gallager, "A Generalized Processor Sharing Approach to Flow Control in Integrated Service Networks: The Multiple Node Case," IEEE/ACM Transaction on Networking, 2 (2), 137-150, 1994.

[3] A. V. Bakre and B.R. Badrinath, "Implementation and Performance Evaluation of Indirect TCP", IEEE Transaction on Computers, 64 (3), pp. 260-278, 1997.

[4] Brian P. Crow, Indra Widjaja, Jeong Geun Kim, Prescott T. Sakai, "IEEE 802.11 Wireless Local Area Networks," IEEE Communications Magazine, Vol. 35, No. 9, September 1997.

[5] Christian Huitema, "Routing in the Internet," Prentice Hall, New Jersey, 1995.

[6] C. Perkins, O. Hodson, and V. Hardman, "A Survey of Packet Loss Recovery Techniques for Streaming Audio," IEEE Network, 12 (5) 1998.

[7] D. M. Chiu and R. Jain, "Analysis of the Increase and Decrease Algorihms for Congestion Avoidance in Computer Networks" Computer Networks and ISDN Systems, 17:1-14, 1989.

[8] David E. McDysan and Darren L. Spohn, "ATM: Theory and Application," McGraw-Hill, New York, 1995.

[9] Daniel O. Awduche, Joe Malcolm, Johnson Agogbua, Mike O'Dell, Jim McManus, "Requirements for Traffic Engineering Over MPLS," Internet draft (work in progress), June 1999.

[10] D. Lin and R. Morris, "Dynamics of Random Early Detection," ACM SIGCOMM'97.

[11] D. Lin, and H. T. Kung, "TCP Fast Recovery Strategies: Analysis and Improvements," Proceedings of IEEE INFOCOM'98, pp. 263-271.

[12] D. Clark and W. Fang, "Explicit Allocation of Best-Effort Packet Delivery Service," IEEE/ACM Transactions on Networking 6 (4), 1998.

[13]  David Eckhardt and Peter Steenkiste, Improving Wireless LAN Performance via Adaptive Local Error Control, Sixth IEEE International Conference on Network Protocols (ICNP'98), Austin, October 1998.

[14]  Duchamp, D. and Reynolds, N. F. "Measured Performance of a Wireless LAN", 17th conference on Local Computer Networks, IEEE 1992, pp. 494-499.

[15]  D. Sisalem and H. Schulzrinne, "The Loss-Delay Adjustment Algorithm: A TCP-friendly Adaptation Scheme," Network and Operating System Support for Digital Audio and Video (NOSSDAV), Cambridge, UK, July 8-10, 1998.

[16]  E. Ayanoglu, S. Paul, T.F. Laportaa, K.K. Sabani and R.D. Gitlin, "AIRMAIL: A Link-Layer Protocol for Wireless Networks," ACM ACM/Baltzer Wireless Networks Journal, 1:47-60, February 1995.

[17]  FreeBSD web site, http://www.freebsd.org.

[18]  F. Bonomi and K. Fendick, "The Rate-Based Flow Control Framework for the Available Bit Rate ATM Service," IEEE Network Magazine, Vol.9, No.2, March/April 1995.

[19]  Go Hasegawa, Masayuki Murata, Hideo Miyahara, "Fairness and Stability of Congestion Control Mechanism of TCP," IEEE INFOCOM'99.

[20]  Gary R. Wright and W. Richard Stevens, "TCP/IP Illustrated, Vol. 2, The Implementation", Addison-Wesley, 1995.

[21]  Gilbert Held, "Frame Relay Networking," John Wiley, 1999.

[22]  G. Carle and E. W. Biersack, "Survey on Error Recovery for IP-based Audio-Visual Multicast Application," IEEE Network Magazine, 1997.

[23]  H. T. Kung and S. Y. Wang, "The Behavior of Competing TCP Connections on a Packet-Switched Ring : A Study Using the Harvard TCP/IP Network Simulator", PDPTA'99 (International Conference on Parallel and Distributed Processing Techniques and Applications), June 28 - July 1, 1999, Las Vegas, USA.

[24]  H. T. Kung and S. Y. Wang, "TCP Trunking: Design, Implementation, and Performance," IEEE ICNP'99.

[25]  H. T. Kung and R. Morris, "Credit-Based Flow Control for ATM Networks," IEEE Network, Vol. 9, No. 2, March/April 1995.

[26] H. Balakrishnan, V. N. Padmanabhanm S. Seshan, and R.H. Katz, "A Comparison of Mechanisms for Improving TCP Performance over Wireless Links", IEEE/ACM Transactions on Networking, December 1997.

[27] H. Balakrishnan, S. Seshan, E. Amir, R.H. Katz, "Improving TCP/IP Performance over Wireless Networks," ACM MOBICOM'95.

[28] H. Balakrishnan and Randy H. Katz, "Explicit Loss Notification and Wireless Web Performance," IEEE Globecom Internet Mini-Conference, Sydney, Australia, November 1998.

[29] H. Balakrishnan, Venkata N. Padmanabhan, Srinivasan Seshan, Mark Stemm, and Randy H. Katz, "TCP Behavior of a Busy Internet Server: Analysis and Improvements," IEEE INFOCOM'98.

[30] Hari Balakrishnan and Hariharan Rahul, "An Integrated Congestion Management Architecture for Internet Hosts," ACM SIGCOMM'99.

[31] Injong Rhee, Nallathambi Balaguru, Goerge Rouskas, "MTCP: Scalable TCP-like Congestion Control for Reliable Multicast," IEEE INFOCOM'99.

[32] J. Postel, "Transmission Control Protocol," RFC 793, September 1981.

[33] J. C. Hoe, "Improving the Start-up Behavior of a Congestion Control Scheme for TCP", SIGCOMM'96, ACM, California, USA, pp. 270-280, 1996.

[34] J. Padhye, V. Firoiu, D. Towsley, and J. Kurose, "Modeling TCP Throughput: A Simple Model and its Empirical Validation", ACM SIGCOMM'98.

[35] J. Touch, S. Ostermann, D. Glover, M. Allman, J. Heidemann, S. Dawkins, J. Semke, K. Scott, J. Griner, D. Tran, T. Henderson, and H. Kruse, "Ongoing TCP Research Related to Satellites," Internet draft, June 1999.

[36] J. Mahdavi and S. Floyd, "TCP-Friendly Unicast Rate-Based Flow Control,", the end2end-interest mailing list, January 8, 1997.

[37] J. Padhye, J. Kurose, D. Towsley and R. Koodli, "A Model Based TCP-Friendly Rate Control Protocol," Network and Operating System Support for Digital Audio and Video (NOSSDAV), June, 1999.

[38] J. Bolot and T. Turletti, "Experience with Control Mechanisms for Packet Video in the Internet," Computer Communications Review, 28 (1), 1998.

[39] J-C Bolot and T. Turletti, "Experience with Control Mechanisms for Packet Video in the Internet," ACM Computer Communication Review, Vol.28 No. 1, 1998.

[40] K. Fall and S. Floyd, "Simulation-based Comparisons of Tahoe, Reno, and SACK TCP", ACM Computer Communication Review, 26(3), pp. 5-21, 1996.

[41] Lorenzo Vicisano, Luigi Rizzo, Jon Crowcroft, "TCP-like Congestion Control for Layered Multicast Data Transfer," IEEE INFOCOM'98.

[42] M. Mathis and J. Mahdavi, "Forward Acknowledgment: Refining TCP Congestion Control", SIGCOMM'96, ACM, California, USA, pp. 281-291, 1996.

[43] M. Mathis, J. Mahdavi, S. Floyd, and A. Romanow, "TCP Selective Acknowledgment Options," RFC 2018, October 1996.

[44] M. Mathis, J. Semke, J. Mahdavi, T. Ott, "The Macroscopic Behavior of the TCP Congestion Avoidance Algorithm",Computer Communication Review, volume 27, number3, July 1997.

[45] M. Allman, D. Glover, L. Sanchez, "Enhancing TCP Over Satellite Channels using Standard Mechanisms," RFC 2488, January 1999.

[46] Mikael Degermark, Mathias Engan, Björn Nordgren, Stephen Pink, "Low-Loss TCP/IP Header Compression for Wireless Networks," ACM MOBICOM'96.

[47] N. Samaraweera and G. Fairhurst, "Reinforcement of TCP/IP Error Recovery for Wireless Communications," Computer Communications Review, 28 (2), 1998.

[48] R. Morris, "TCP Behavior with Many Flows," IEEE ICNP'97.

[49] R. Bruyeron, B. Hemon, and L. Zhang, "Experiments with TCP Selective Acknowledgment," ACM Communication Review, Vol. 28, No. 2, April 1998.

[50] Radia Perlman, "Interconncetions: Bridges amd Routers," Addison-Wesley, 1992.

[51] R. Callon, N. Feldman, A. Fredette, G. Swallow, A. Viswanathan, "A Framework for Multiprotocol Label Switching,", Internet draft (work in progress), June 1999

[52] Reza Rejaie, Mark Handley, Deborah Estrin, "An End-to-end Rate-based Congestion Control Mechanism for Realtime Streams in the Internet," IEEE INFOCOM'99.

[53] S. Floyd, "TCP and Explicit Congestion Notification," ACM Computer Communication Review, 24 (5), 1994, pp. 10-23.

[54] S. Floyd and V. Jacobson, "Random Early Detection Gateways for Congestion Avoidance", IEEE/ACM Transactions on Networking, 1 (4), 1993, pp. 397-413.

[55] S. Floyd, "Connections with Multiple Congested Gateways in Packet-Switched Networks Part 1: One-way traffic," Computer Communications Review, 21 (5), 1991.

[56] S. Floyd and K. Fall, "Promoting the Use of End-to-End Congestion Control in the Internet," IEEE/ACM Transactions on Networking, August 1999

[57] S. Blake, D. Black, M. Carlson, E. Davies, Z. Wang, W. Weiss, "An Architecture for Differentiated Services,", RFC 2475, December 1998.

[58] S. Y. Wang and H.T. Kung, "A Simple Methodology for Constructing an Extensible and High-Fidelity TCP/IP Network Simulator," INFOCOM'99.

[59] V. Jacobson, "Congestion Avoidance and Control," ACM SIGCOMM'88, pp. 314-329, 1988.

[60] V. Jacobson, "Compressing TCP/IP Headers for Low-Speed Serial Links," RFC 1144.

[61] V. N. Padmanabhan and J.C. Mogul. "Improving HTTP Latency," The second International WWW Conference," October 1994.

[62] W. R. Stevens, "TCP/IP Illustrated: The protocols, Vol:1". Addison Wesley, New York, 1994.

[63] W. R. Stevens, "TCP Slow Start, Congestion Avoidance, Fast Retransmission, and Fast Recovery Algorithms," RFC 2001, January 1997.

[64] W. R. Stevens, "TCP Slow Start, Congestion Avoidance, Fast Retransmit, and Fast Recovery Algorithms," RFC 2001, January 1997.

[65] WaveLAN web site, http://www.wavelan.com.